# Combined CO & Dust Scaling Relations of Depletion Time and Molecular Gas Fractions with Cosmic Time, Specific Star Formation Rate and Stellar Mass[1]

R.Genzel[1,2,3], L.J.Tacconi[1], D.Lutz[1], A. Saintonge[4], S.Berta[1], B.Magnelli[5], F.Combes[6], S.García-Burillo[7], R.Neri[8], A.Bolatto[9], T.Contini[10], S.Lilly[11], J.Boissier[8], F.Boone[10], N.Bouché[10], F.Bournaud[12], A.Burkert[13,1], M.Carollo[11], L.Colina[14], M.C.Cooper[15], P.Cox[16], C.Feruglio[8], N.M. Förster Schreiber[1], J.Freundlich[6], J.Gracia-Carpio[1], S.Juneau[12], K.Kovac[11], M.Lippa[1], T.Naab[17], P.Salome[6], A.Renzini[18], A.Sternberg[19], F.Walter[20], B.Weiner[21], A.Weiss[22] & S.Wuyts[1]

[1] Max-Planck-Institut für extraterrestrische Physik (MPE), Giessenbachstr., 85748 Garching, FRG
( linda@mpe.mpg.de, genzel@mpe.mpg.de )
[2] Dept. of Physics, Le Conte Hall, University of California, 94720 Berkeley, USA
[3] Dept. of Astronomy, Campbell Hall, University of California, Berkeley, CA 94720, USA
[4] Department of Physics & Astronomy, University College London, Gower Place, London WC1E 6BT, UK
[5] Argelander-Institut für Astronomie, Universität Bonn, Auf dem Hügel 71, 53121 Bonn, FRG
[6] Observatoire de Paris, LERMA, CNRS, 61 Av. de l'Observatoire, F-75014 Paris, FR
[7] Observatorio Astronómico Nacional-OAN, Observatorio de Madrid, Alfonso XII, 3, 28014 - Madrid, SP
[8] IRAM, 300 Rue de la Piscine, 38406 St.Martin d'Heres, Grenoble, France
[9] Dept. of Astronomy, University of Maryland, College Park, MD 20742-2421, USA
[10] Institut d'Astrophysique et de Planétologie, Universite de Toulouse, 9 Avenue du Colonel Roche BP 44346 - 31028 Toulouse Cedex 4, FR
[11] Institute of Astronomy, Department of Physics, Eidgenössische Technische Hochschule, ETH Zürich, CH-8093, SW
[12] Service d'Astrophysique, DAPNIA, CEA/Saclay, F-91191 Gif-sur-Yvette Cedex, FR
[13] Universitätssternwarte der Ludwig-Maximiliansuniversität, Scheinerstr. 1, D-81679 München, FRG
[14] CSIC Instituto Estructura Materia, C/ Serrano 121, 28006 Madrid, SP
[15] Dept. of Physics & Astronomy, Frederick Reines Hall, University of California, Irvine, CA 92697
[16] ALMA Santiago Central Office, Alonso de Córdova 3107, Vitacura, Santiago, CH
[17] Max-Planck Institut für Astrophysik, Karl Schwarzschildstrasse 1, D-85748 Garching, FRG
[18] Osservatorio Astronomico di Padova, Vicolo dell'Osservatorio 5, Padova, I-35122, IT
[19] School of Physics and Astronomy, Tel Aviv University, Tel Aviv 69978, Israel
[20] Max Planck Institut für Astronomie (MPIA), Königstuhl 17, 69117 Heidelberg, FRG
[21] Steward Observatory, 933 N. Cherry Ave., University of Arizona, Tucson AZ 85721-0065, USA
[22] Max Planck Institut für Radioastronomie (MPIfR), Auf dem Hügel 69, 53121 Bonn, FRG

[1] Based on observations with the Plateau de Bure millimetre interferometer, operated by the Institute for Radio Astronomy in the Millimetre Range (IRAM), which is funded by a partnership of INSU/CNRS (France), MPG (Germany) and IGN (Spain).

# ABSTRACT

We combine molecular gas masses inferred from CO emission in 500 star forming galaxies (SFGs) between z=0 and 3, from the IRAM-COLDGASS, PHIBSS1/2 and other surveys, with gas masses derived from Herschel far-IR dust measurements in 512 galaxy stacks over the same stellar mass/redshift range. We constrain the scaling relations of molecular gas depletion time scale ($t_{depl}$) and gas to stellar mass ratio ($M_{molgas}/M_*$) of SFGs near the star formation 'main-sequence' with redshift, specific star formation rate (*sSFR*) and stellar mass ($M_*$). The CO- and dust-based scaling relations agree remarkably well. This suggests that the CO → $H_2$ mass conversion factor varies little within ±0.6dex of the main sequence (*sSFR(ms,z,M*)*), and less than 0.3dex throughout this redshift range. This study builds on and strengthens the results of earlier work. We find that $t_{depl}$ scales as $(1+z)^{-0.3} \times (sSFR/sSFR(ms,z,M_*))^{-0.5}$, with little dependence on $M_*$. The resulting steep redshift dependence of $M_{molgas}/M_* \propto (1+z)^3$ mirrors that of the *sSFR* and probably reflects the gas supply rate. The decreasing gas fractions at high $M_*$ are driven by the flattening of the *SFR-M*$_*$ relation. Throughout the redshift range probed a larger *sSFR* at constant $M_*$ is due to a combination of an increasing gas fraction and a decreasing depletion time scale. As a result galaxy integrated samples of the $M_{molgas}$-*SFR* rate relation exhibit a super-linear slope, which increases with the range of *sSFR*. With these new relations it is now possible to determine $M_{molgas}$ with an accuracy of ±0.1dex in relative terms, and ±0.2dex including systematic uncertainties.

# 1. Introduction

   Stars form from dusty, molecular interstellar gas (McKee & Ostriker 2007, Kennicutt & Evans 2012). In the Milky Way and nearby galaxies arguably all star formation occurs in massive ($10^4...10^{6.5}$ $M_\odot$), dusty and dense (n($H_2$)~$10^2...10^5$ cm$^{-3}$), cold ($T_{gas}$~10-40 K), 'giant molecular clouds' (GMCs) that are near or in virial equilibrium (Solomon et al. 1987, Bolatto et al. 2008, McKee & Ostriker 2007, but see Dobbs, Burkert & Pringle 2011, Dobbs & Pringle 2013). The star formation rates on galactic scales or star formation surface densities on sub-galactic scales down to a few kpc are empirically most strongly correlated with molecular gas (or dust) masses, or surface densities, while there is little or no correlation between star formation and neutral atomic hydrogen (Kennicutt 1989, Kennicutt et al. 2007, Bigiel et al. 2008, 2011, Leroy et al. 2008, 2013, Schruba et al. 2011). However, it is not clear whether high molecular content as such is causally required for the onset of star formation (Glover & Clark 2012). Rather the key ingredients may be the combination of high gas volume density and sufficient dust shielding ($A_V$>7, $\Sigma_{gas}$>100 $M_\odot pc^{-2}$) to decouple the dense cores from the external radiation field and allow it to cool and initiate collapse; these conditions may then also be conducive to molecule formation (Glover & Clark 2012, Krumholz, Leroy & McKee 2011, Heiderman et al.2010, Lada et al. 2012).

   About 90% of the cosmic star formation between z=0 and 2.5 occurs in galaxies that lie near the so-called '***star formation main sequence***' (Rodighiero et al. 2011, Sargent et al. 2012), which is a fairly tight (±0.3 dex scatter), near-linear relationship between stellar mass and star formation rate (Schiminovich et al. 2007, Noeske et al. 2007, Elbaz et al. 2007, 2011, Daddi et al. 2007, Panella et al. 2009, Peng et al. 2010, Rodighiero et al. 2010, Karim et al. 2011, Salmi et al. 2012, Whitaker et al. 2012, 2014, Lilly et al. 2013). From the NEWFIRM medium band survey in the AEGIS and COSMOS fields Whitaker et al. (2012) have proposed an analytic fitting function of the center line of this sequence as a function of redshift (0<z<2.5) and stellar mass (for $M_* \geq 10^{10}$ $M_\odot$)

$$\log(sSFR(ms, z, M_*)) = -1.12 + 1.14z - 0.19z^2 - (0.3 + 0.13z) \times (\log M_* - 10.5) \quad (Gyr^{-1}) \quad (1),$$

where the specific star formation rate *sSFR* (Gyr$^{-1}$) is the ratio of star formation rate *SFR* ($M_\odot yr^{-1}$) and stellar mass $M_*$ ($M_\odot$).

   Main-sequence SFGs are characterized by disky, exponential rest-UV/rest-optical light distributions ($n_{Sersic}$~1-2, Wuyts et al. 2011b) and a strong majority is rotation dominated (e.g. Shapiro et al. 2008, Förster Schreiber et al. 2009, Newman et al. 2013, Wisnioski et al. 2014). The tightness and time independent shape of the main sequence suggests that star forming galaxies grow along the sequence in an equilibrium of gas accretion, star formation and gas outflows (the 'gas regulator model': Bouché et al. 2010, Davé et al. 2012, Lilly et al. 2013, Peng & Maiolino 2014). At z>1 main-sequence SFGs double their mass on a typical time scale of ~500 Myrs but their growth appears to halt suddenly when they reach the Schechter mass, $M_* \sim 10^{10.8..11}$ $M_\odot$ (Conroy & Wechsler 2009, Peng et al. 2010). For a better understanding of the origin and evolution of this equilibrium evolution of the main sequence population, the goal of current studies is to establish how (efficiently) the conversion from cool gas to stars proceeds on a global galactic scale, and how this efficiency and the galaxies' gas reservoirs change as a function of cosmic epoch (redshift), stellar mass, star formation rate, galaxy size/internal structure, gas motions and environmental parameters (see discussions in Daddi et al. 2010a,b, Tacconi et al. 2010, 2013, Genzel et al. 2010, Bouché et al. 2010, Lilly et al. 2013, Davé et al. 2011, 1012, Lagos et al. 2011, Fu et al. 2012).

Motivated by the growing body of recent evidence in the literature on the physical properties of main-sequence galaxy populations as a function of z (da Cunha et al. 2010, Elbaz et al. 2011, Gracia-Carpio et al. 2011, Wuyts et al. 2011, Magdis et al. 2012b, Nordon et al. 2012, Saintonge et al. 2012, Tacconi et al. 2013, Magnelli et al. 2014), our tenet in this paper is *that the scaling relations depend primarily on the location of a galaxy relative to the main-sequence line (sSFR/sSFR(ms,z,M∗)), and only indirectly on the absolute value of the sSFR*.

The parameterization of the star formation main sequence as a function of redshift and stellar mass varies among the different studies mentioned above. This can be understood by different sample selections, survey completeness, methodology applied to derive $M_*$ and *SFR*s, among other factors. Perhaps most importantly, the inferred slope of the main-sequence as a function of $M_*$ depends on whether the sample is mass selected (including quenched galaxies leading to a steep slope, $d(sSFR)/d(logM_*) = -0.3..-0.5$), or UV/optical magnitude-color selected (selecting mainly star forming galaxies, shallow slope, $d(sSFR)/d(logM_*) = -0.1..0$). The Whitaker et al. (2012) fits (see also Whitaker et al. 2014) provide a good representation of the **actual locus** of the near-main-sequence SFGs in our samples above $\log(M_*/M_\odot) \sim 10..10.2$. Their selection on the basis of stellar mass and restframe UVJ colors includes also red and dusty star forming galaxies. In contrast a main-sequence with $d(sSFR)/d(logM_*) \sim 0$ would be the expected slope of actively star forming galaxies growing in the equilibrium gas regulator framework (Lilly et al. 2013). The fact that at high stellar masses the slope of the main-sequence seems to steepen would then mean that the most massive star forming galaxies are beginning to drop below this ideal line and quench. We discuss in section 4.2 the impact of different parameterizations of the main sequence relation on the scaling relations.

To determine and quantify these dependencies, it is convenient to determine first the gas depletion time scale, $t_{dep}$, as a function of the above mentioned parameters

$$t_{dep} = M_{gas}/SFR \quad \text{or,}$$
$$t_{dep} = \Sigma_{gas}/\Sigma_{SFR} \quad (2),$$

where $M_{gas}$ and $\Sigma_{gas}$ are the gas mass and surface density, *SFR* and $\Sigma_{SFR}$ the total rate and surface density of star formation (the 'Kennicutt-Schmidt' relation between gas and star formation rate, Kennicutt 1998). The first equation is appropriate for galaxy integrated, and the second for spatially resolved data.

Given the discussion above, it is most appropriate to concentrate here on the molecular gas depletion time scale, where the total gas mass and surface density on the right side of the equations in (2) are replaced by the molecular hydrogen mass and surface density, including the standard correction for helium (~36% in mass), and for the photo-dissociated surface layers of the molecular clouds that are fully molecular in $H_2$ but 'dark' (i.e. dissociated) in CO (Wolfire, Hollenbach & McKee 2010, Bolatto, Wolfire & Leroy 2013).

The virtue of the empirical depletion time scale (without any reference to its physical interpretation) is that it is easily accessible to global measurements of the standard tracers of star formation and gas (i.e. stellar and infrared luminosity, CO 1-0, 2-1, 3-2 line luminosity, HI mass, dust mass) in a large number of galaxies (e.g. Young & Scoville 1991, Solomon & Sage 1988, Gao & Solomon 2004, Scoville 2013). In the recent IRAM COLDGASS survey Saintonge et al. (2011a,b, 2012) have observed the galaxy integrated CO 1-0 line flux in 365 mass selected ($M_*>10^{10}\,M_\odot$) SDSS galaxies between z=0.025…0.05. This homogeneously calibrated, purely mass selected survey can be directly connected to the properties of the overall SDSS parent sample. Saintonge et al. (2011b, 2012) find an average depletion time of about 1.2 Gyr for galaxies near the star formation main sequence (Brinchmann et al. 2004, Schiminovich et al. 2007), but a decrease in the depletion time above, and an increase in the depletion time scale below the main sequence, toward the sequence of passive galaxies. In the IRAM HERACLES survey Bigiel et al. (2008, 2011), Leroy et al. (2008, 2013) and Schruba et al. (2011) studied the spatial distribution of CO 2-1 emission on subgalactic scales (resolution ~1 kpc) in 30 local disk and dwarf star forming galaxies near the main sequence. They find a relatively constant depletion time scale of about 2.2 Gyrs.[2]

Once the depletion time scale is determined, baryonic molecular gas mass fractions can then be computed in a straightforward manner from

$$\frac{M_{molgas}}{M_*} = \frac{M_{molgas}}{SFR} \times \frac{SFR}{M_*} = t_{dep} \times sSFR \quad ,$$

$$\text{and } f_{molgas} = \frac{M_{molgas}}{M_{molgas}+M_*} \quad (3).$$

---

[2] the factor 2 (0.3 dex) difference in the depletion times inferred from the COLDGASS (Saintonge et al. 2011) and HERACLES (Bigiel et al. 2008, Leroy et al. 2013) surveys owes to the combination of different computation of star formation rates, SED modeling in the former, and from UV+mid-IR or Hα+mid-IR in the latter (~30% effect), and the weighting scheme of different data points, integration over the entire galaxy in the former, and averaging individual line of sights with CO detections in the latter, including the treatment of diffuse Hα/IR emission (~60% effect). This difference is well understood but might be taken as an estimate of the underlying systematic uncertainties. The calibration and methodology of the high-z data discussed in this paper is close to that of the COLDGASS survey approach, although for most galaxies in the PHIBSS1&2 surveys star formation rates are cross-calibrated to the UV+mid/far-IR scale through a "ladder" approach (Wuyts et al. 2011a).

Until a few years ago, studies of the gas content in z>0.5 galaxies were restricted to luminous, gas and dust rich, outliers, such as starbursts and mergers, significantly above the main-sequence line at their respective redshifts (e.g. Greve et al. 2005, Tacconi et al. 2006, 2008, Riechers 2013, Carilli & Walter 2013, Bothwell et al. 2013). With the availability of more sensitive receivers at the IRAM Plateau de Bure mm-interferometer (PdBI: Guilloteau et al.1992, Cox 2011, Tacconi et al. 2010, 2013, Daddi et al. 2008, 2010a), the start of the science phase of ALMA, and the availability of dust observations from the Herschel PACS and SPIRE instruments, this situation has started to change dramatically and rapidly. Nevertheless it is, and will be for the foreseeable future, unrealistic to expect that one can carry out direct (molecular) gas mass estimates for galaxy sample sizes approaching or comparable to those in the panoramic UV, optical/near-IR and mid-IR/far-IR surveys ($10^{4...5.5}$ galaxies in the standard "cosmological" fields).

In the present paper we instead use the presently available data on star forming galaxies near and above the main sequence from the current epoch (z~0) to the peak of the cosmic star formation activity (z~1-3) to determine how the molecular depletion times (and gas fractions) vary with redshift, star formation rate and stellar mass. With scaling relations in hand, it is then possible to predict the molecular gas properties of larger samples just on the basis of these basic input parameters. We take advantage of the availability of both CO-based and dust-based molecular gas mass determinations over the same range in parameters to compare these independent methods, and in particular, establish, reliable 'zero points'.

Throughout, we adopt a Chabrier (2003) stellar initial mass function and a $\Lambda$CDM cosmology with $H_0 = 70$ km s$^{-1}$ and $\Omega_m = 0.3$.

# 2. Observations

### 2.1 CO observations

To explore the cold molecular gas in SFGs covering the entire redshift range from z=0 to 4, the stellar mass range of $M_*=10^{9.8}$ to $10^{11.8}$ $M_\odot$, and at a given redshift and stellar mass, star formation rates from about $10^{-1}$ to $10^2$ times the main-sequence star formation rate, we collected 500 CO detections of star forming galaxies near, below and above the main sequence from a number of concurrent molecular surveys with CO 1-0, 2-1, 3-2 (and in two cases 4-3) rotational line emission (Table 1). We include:

1. 216 detections and 1 stack detection (much below the main-sequence) of CO 1-0 emission above and below the main sequence between z=0.025-0.05 from the final COLDGASS survey with the IRAM 30m telescope (Saintonge et al. 2011a,b, 2014 in prep.). We note that the star formation rates in that survey have been updated from earlier UV-/optical SED fitting (Saintonge et al. 2011a) with mid-IR star formation rates from WISE, (Saintonge et al. in prep., Huang & Kauffmann 2014);
2. 90 CO 1-0 detections with the IRAM 30m of z=0.002-0.09 luminous and ultra-luminous IR-galaxies (LIRGs and ULIRGs) from the GOALS survey (Armus et al. 2009), from the work of Gao & Solomon (2004), Gracia-Carpio et al. (2008, 2009, and priv. comm.), and Garcia-Burillo et al. (2012);
3. 31 CO 1-0 or 3-2 detections of (above main-sequence) SFGs between z=0.06 and 0.5 with the CARMA millimeter array from the EGNOG survey (Bauermeister et al. 2013);
4. 14 CO 2-1 or 3-2 detections at z=0.6-0.9 and 18 CO 1-0 detections at z=0.2-0.58 (significantly above-main-sequence) ULIRGs with the IRAM 30m telescope from Combes et al. (2011,2013);
5. 11 CO 2-1 or 3-2 detections of near main-sequence SFGs between z=0.5 and 3.2 from Daddi et al. (2010a) and Magdis et al. (2012a), obtained with the IRAM PdBI;
6. 6 CO 2-1 detections of z=1-1.2 main-sequence SFGs selected from the Herschel-PEP survey (Lutz et al. 2011), obtained with the IRAM PdBI (Magnelli et al. 2012a);
7. 52 detections of CO 3-2 emission in main-sequence SFGs in two redshift slices at z=1-1.5 (38) and z=2-2.5 (14) as part of the PHIBSS1 survey with the IRAM PdBI (Tacconi et al. 2010, 2013);
8. 31 detections (and 2 upper limits) of CO 2-1 or 3-2 in main sequence SFGs between z=0.5 and 1, and 3 at z~2, as part of the PHIBSS2 survey with the IRAM PdBI (Tacconi, Combes et al. 2014, in preparation);
9. 19 CO 2-1, 3-2 or 4-3 detections of above main-sequence submillimeter galaxies (SMGs) between z=1.2 and 3.4, obtained with the IRAM PdBI by Greve et al. (2005), Tacconi et al. (2006, 2008) and Bothwell et al. (2013);
10. 8 CO 3-2 detections of z=1.4 to 3.2 lensed main-sequence SFGs obtained with the IRAM PdBI (Saintonge et al. 2013, and references therein).

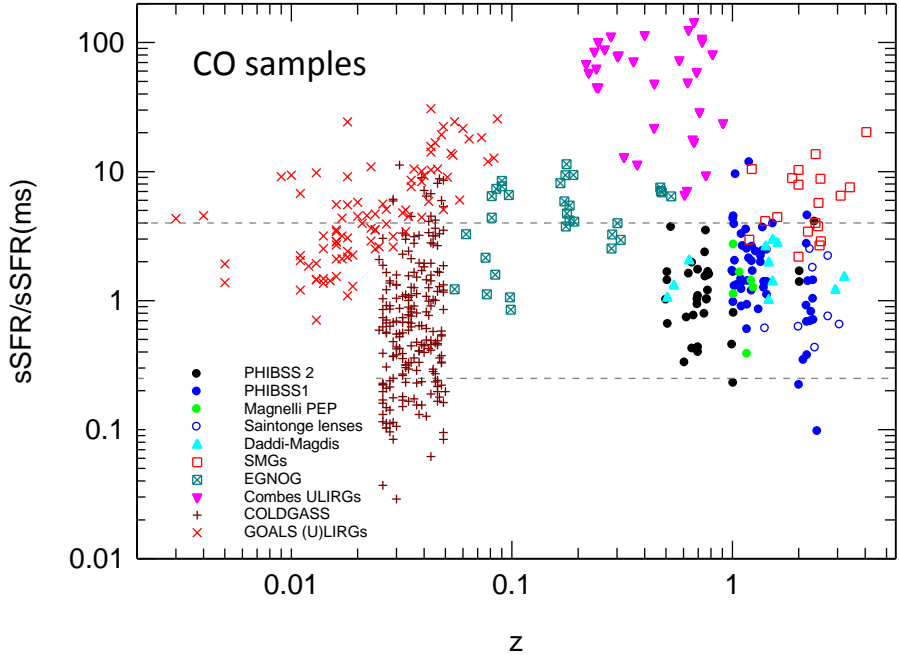

Figure 1: Distribution in the redshift –specific star formation rate plane of the 500 SFGs with integrated CO (1-0, 2-1, 3-2 and 4-3) flux measurements used in this paper. The various symbols denote the different publications from which these measurements were taken, as discussed in the text (section 2.1, Table 1). The vertical axis is normalized so that the mid-line of the star formation sequence at each redshift is at unity, using the scaling relations *sSFR(ms,z,M∗)* from equation (1) (Whitaker et al. 2012). Horizontal dashed lines mark the upper and lower range of the main sequence, ±0.6 dex from that mid-line.

The redshift-*sSFR* coverage of this sample is shown in Figure 1, with the different symbols denoting the various surveys mentioned in our listing above. Owing to the sensitivity limits the overall distribution in *z-sSFR* space is biased to SFGs above the main sequence. However, the more recent extensive surveys at the IRAM telescopes at z~0.03 (COLDGASS), z~0.7 (PHIBSS2), z~1.2 (PHIBSS1+2) and z=2.2 (PHIBSS1) have begun to establish a decent coverage of massive SFGs above and below the main-sequence line. Most of these data are benchmark sub-samples of large UV/optical/infrared/radio imaging surveys with spectroscopic redshifts, and well established and relatively homogeneous stellar and star formation properties. The COLDGASS sample is drawn from SDSS. PHIBSS1+2 and the data of Daddi et al. (2010a), Magdis et al. (2012a) and Magnelli et al. (2012a) are selected from deep rest-frame UV-/optical imaging surveys in EGS (Davis et al. 2007, Newman et al. 2013, Cooper et al. 2012), GOODS N (Giavalisco et al. 2004, Berta et al. 2010) and COSMOS (Scoville et al. 2007, Lilly et al. 2007, 2009), including the recent CANDELS J- and H-band HST imaging (Grogin et al. 2011, Koekemoer et al. 2011) and 3D-HST grism spectroscopy (Brammer et al. 2012, Skelton et al. 2014), as well as D3a (Kong et al. 2006) and the BX/BM samples of Steidel et al. (2004) and Adelberger et al. (2004).

We have binned the 500 SFGs of our CO sample into 6 redshift bins (Table 1). The number of SFGs in each of the five higher z bins is comparable (28-49). The highest four bins have a good coverage of the main-sequence population, while the lowest of these non-local bins (z=0.05-0.45) contains mostly above main sequence, starburst outliers. There are few galaxies significantly below the main sequence, for the obvious reason of detectability. The lowest redshift bin (mostly COLDGASS) naturally contains by far the largest number of galaxies (296 of the 500 galaxies). This imbalance needs to be taken into account carefully when considering the scaling relations.

We emphasize that the majority of our final sample of ~500 galaxies are near-main sequence SFGs $\Delta log(sSFR/sSFR(ms))=\pm 0.6$ (dashed horizontal lines in Figure 1), with a few below main-sequence (mainly from the COLDGASS sample at z=0), and ~130 (26%) above main-sequence starburst outliers. The focus of this paper is on the near-main sequence population.

### *2.1.1 Derivation of molecular gas masses*

Observations of giant molecular clouds (GMCs) in the Milky Way and nearby galaxies have established that the integrated line flux of $^{12}$CO millimeter rotational lines can be used to infer molecular gas masses, although the CO molecule only makes up a small fraction of the entire gas mass, and its lower rotational lines (1-0, 2-1, 3-2) are almost always very optically thick (Dickman, Snell & Schloerb 1986, Solomon et al. 1987, Bolatto et al. 2013). This is because the CO emission comes from moderately dense (volume average densities $<n(H_2)> \sim 200$ cm$^{-3}$, column densities $N(H_2) \sim 10^{22}$ cm$^{-2}$), self-gravitating GMCs of kinetic temperature 10-50 K. Dickman et al. (1986) and Solomon et al. (1987) have shown that in this 'virial' regime, or if the emission comes from an ensemble of similar mass, near-virialized clouds spread in velocity by galactic rotation, the integrated line CO line luminosity $L'_{CO} = \int_{source} \int_{line} T_R \, dv \, dA$ (in K km/s pc$^2$, $T_R$ is the Rayleigh-Jeans source brightness temperature as a function of Doppler velocity v) is proportional to the total gas mass in the cloud/galaxy. In this 'cloud counting' technique the total molecular gas mass (including a 36% mass correction for helium) then depends on the observed CO J →J-1 line flux $F_{CO\,J}$, source luminosity distance $D_L$, redshift z and observed line wavelength $\lambda_{obs\,J} = \lambda_{rest\,J}(1+z)$ as (Solomon et al. 1997)

$$M_{molgas}/M_\odot = \alpha_{CO\,1} \times L'_{CO\,1}$$

$$= 1.75 \times 10^9 \left( \frac{\alpha_{CO\,1} \times R_{1J}}{\alpha_{MW}} \right) \times \left( \frac{F_{CO\,J}}{\text{Jy km/s}} \right) \times (1+z)^{-3} \times \left( \frac{\lambda_{obs\,J}}{\text{mm}} \right)^2 \times \left( \frac{D_L}{\text{Gpc}} \right)^2 \quad (4).$$

Here $\alpha_{CO\,1}$ is an empirical 'conversion factor' to transform the observed quantity (CO

luminosity in the 1-0 transition) to the inferred physical quantity (molecular gas mass), and $R_{1J}$ is the ratio of the 1-0 to the J – (J-1) CO line luminosity, $R_{1J}=L'_{CO\ 1\text{-}0}/L'_{CO\ J-(J\text{-}1)}$.

*From conversion 'factor' to conversion 'function':* From theoretical considerations the CO conversion factor in equation (4) is expected to be a function of several physical parameters (Narayanan et al. 2011, 2012, Feldmann et al. 2012 a,b). In the virial/cloud counting model α depends on the ratio of the square root of the average cloud density $<n(H_2)>$ and the equivalent Rayleigh-Jeans brightness temperature $T_{R\ J}$ of the CO transition J→J-1. It also increases with the inverse of the metallicity $Z$ (see Leroy et al. 2011, Genzel et al. 2012 and Bolatto et al. 2013 for more detailed discussions of the observational evidence),

$$\alpha_{CO\ J} = \zeta \left( \frac{(<n(H_2)>)^{1/2}}{T_{R\ J}} \right) \times \chi(Z) \qquad (5).$$

In the Milky Way and nearby star forming galaxies with near solar metallicity, as well as in dense star forming clumps of lower mass, lower metallicity galaxies, the empirical CO 1-0 conversion factor $\alpha_{CO\ 1}$ determined with dynamical, dust and γ-ray calibrations are broadly consistent with a single value of $\alpha_{CO\ 1} = \alpha_{MW} = 4.36 \pm 0.9$ (M$_\odot$/(K km/s pc$^2$)), equivalent to $X_{CO}=N(H_2)/(T_{RJ=1}\Delta v)= 2\times10^{20}$ (cm$^{-2}$/(K km/s), Strong & Mattox 1996, Dame, Hartmann & Thaddeus 2001, Grenier, Casandijan & Terrier 2005, Bolatto et al. 2008, Leroy et al. 2011, Abdo et al. 2010, Ostriker, McKee & Leroy 2010, Bolatto et al. 2013).

*Metallicity dependence of the conversion factor*: For galaxies of sub-solar gas phase metallicity, the conversion factor and metallicity are inversely correlated, as the result of an increasing fraction of the molecular hydrogen gas column that is photo-dissociated, resulting in molecular gas that is deficient ('dark') in CO (Wilson 1995, Arimoto, Sofue & Tsujimoto 1996, Israel 2000, Wolfire et al. 2010, Leroy et al. 2011, Genzel et al. 2012, Bolatto et al. 2013, Sternberg et al. 2014). Motivated by the theoretical work of Wolfire et al. (2010) on the photo-dissociation of clouds with a range of hydrogen densities and UV radiation field intensities, but with a constant hydrogen column, Bolatto et al. (2013) have proposed the following fitting function for χ(Z),

$$\chi(Z) = 0.67 \times \exp(0.36 \times 10^{-(12+\log(O/H)-8.67)}) \qquad (6),$$

where 12+log(O/H) is the gas phase oxygen abundance in the galaxy on the Pettini & Pagel (2004) calibration scale, with the solar abundance of 8.67 (Asplund et al. 2004). Equation (6) assumes an average GMC hydrogen column density of 100 M$_\odot$ pc$^{-2}$, or $9\times10^{21}$ cm$^{-2}$. Genzel et al. (2012) have combined the local (Leroy et al. 2011) and high-z empirical evidence for a second fitting function,

$$\chi(Z) = 10^{-1.27\times(12+\log(O/H)-8.67)} \qquad (7).$$

For the near-solar metallicities typical for most SFGs in our overall sample (96% of the 1012 SFGs are between 12+log(O/H)=8.55 and 8.75 on the PP04 scale), equations (6) and (7) yield values for χ(Z) within ±0.12 dex of each other. We thus took the geometric mean of (6) and (7) in estimating the gas masses from CO in this paper. Note that this approach is not applicable for significantly sub-solar metallicity galaxies. Between 12+log(O/H) = 7.9 and 8.4 equation (7) implies a correction 0.22 to 0.32 dex greater than equation (6).

*CO ladder excitation dependence of the conversion factor:* To convert the CO 2-1 and 3-2 luminosities in the near main-sequence SFGs (at all redshifts) to an equivalent CO 1-0 luminosity we apply a correction factor of $R_{1J}$ = 1.3 and 2 to correct for the lower Rayleigh-Jeans brightness temperature of the J – (J-1) relative to the 1-0 transition. This 'excitation' correction entails a combination of the Planck correction (for a finite rotational temperature), as well as a correction for a sub-thermal population in the upper rotational levels. For above main sequence SMGs and ULIRGs (*sSFR/sSFR(ms,z,M*)*>4) we take $R_{1J}$= 1.2, 1.9 and 2.4 for the 2-1, 3-2 and 4-3 transitions. These correction factors are empirically motivated by recent CO ladder observations in low- and high-z SFGs (Weiss et al. 2007, Dannerbauer et al. 2009, Ivison et al. 2011, Riechers et al. 2010, Combes et al. 2013, Bauermeister et al. 2013, Bothwell et al. 2013, Aravena et al. 2014, Daddi et al. 2014). While these correction factors undoubtedly vary from galaxy to galaxy, their scatter is unlikely to be greater than ±0.1 dex, as judged from the recent data sets.

*Density-temperature dependence of the conversion factor:* This leaves the correction factor/function $\zeta\left(\frac{(<n(H_2)>)^{1/2}}{T_{RJ}}\right)$, which is correlated with the star formation rate at a given mass and redshift, that is, the vertical location in the stellar mass – star formation rate plane (Elbaz et al. 2011, Gracia-Carpio et al. 2011, Nordon et al. 2012, Lada et al. 2012). Our initial assumption is that this function is a constant of order unity. As we show below (section 4.1), this assumption is justified for the z=0…3 near-main sequence population, which is at the focus of this paper. However, this assumption is very likely not appropriate for outliers/starbursts high above the main-sequence (Daddi et al. 2010b, Genzel et al. 2010 (and references therein), Magdis et al. 2012b, Sargent et al. 2012, 2014, Tan et al. 2014). For instance, for extreme z=0 ULIRGs there is good evidence from dynamical arguments that $α_{CO}$ is 0.8 to 1.5, or 0.46 to 0.74 dex smaller than the Milky Way value, perhaps implying the presence of a second, 'starburst' star formation mode with ~5 times greater star formation efficiency (Scoville et al. 1997, Downes & Solomon 1998). Daddi et al. (2010b), Genzel et al. (2010) and Sargent et al. (2014) have incorporated this information for the input assumptions of equation (5). However, this approach has since come under criticism (e.g. Ostriker & Shetty 2011, Kennicutt & Evans 2012, Krumholz, Dekel & McKee 2012) on the grounds that the resulting smaller depletion time scales then immediately introduce a bi-modal, and thus probably unphysical gas-star formation relation. We take the approach in this paper of not including a priori such a correction factor for the outlier-population, and then derive in section 4.1 quantitative constraints on the scaling of $α_{CO}$ with *sSFR/sSFR(ms,z,M*)* from the comparison of the CO-data to the dust data (not affected by the conversion factor issue). From this comparison we will show that for the near-main population that is the focus of this paper, this simplified approach delivers a good description of the conversion factor.

Our starting point in this paper thus is to use for all 500 SFGs

$$\alpha_{0J} = \alpha_{MW} \times \chi(Z) \times R_{1J} \quad (8)$$

to derive molecular gas masses from CO observations.

## 2.2 Dust observations

As part of the Herschel-PEP (Lutz et al. 2011) and Herschel-HERMES (Oliver et al. 2012) far-IR continuum surveys, Magnelli et al. (2014) have established 100 to 500µm far-IR SEDs from stacking PACS and SPIRE photometry in 8846, 4753 and 254 749 K- and I-selected SFGs in the GOODS-N, GOODS-S and COSMOS fields, respectively. For details of the methodology we refer to Magnelli et al. (2014). Briefly, star formation rates are calibrated onto the Wuyts et al. (2011a) ladder of UV-, mid-IR and far-IR based indicators. Since the far-IR detection rate drops with increasing redshift and decreasing SFR and $M_*$, it is necessary to average many individual data points to determine good far-IR SEDs as a function of z, SFR and $M_*$. For this purpose, Magnelli et al. (2014) binned the data onto a three dimensional grid in *z*, *SFR* and *$M_*$* and then stacked the photometry in each bin. Next Magnelli et al. determined for each resulting SED the dust temperature, by fitting to model SEDs from the library of Dale and Helou (2002), for which dust temperatures were established from single optically thin, modified blackbody fits with emission index β=1.5 (Table A.1 in Magnelli et al. 2014). To ensure constrained SED shapes, we use only bins where the stacked photometry is detected at >3σ in at least 3 bands that encompass the fitted SED peak, and the reduced $\chi^2$ of the fit is less than 2. In the median, detections in our stacked SEDs reach out to a rest wavelength of 223µm, and only ~10% stop at ≤160µm.

From these stacks we computed dust masses using Draine & Li (2007) models. We fitted the models following the procedure prescribed by these authors, and widely used in the literature, adopting the Li & Draine (2001) values of dust opacity as a function of wavelength. We limited the parameter space to the range suggested by Draine et al. (2007) for galaxies missing sub-mm data, based on their analysis of local SINGS galaxies. Draine et al. (2007) compared dust masses of local SINGS galaxies obtained using SEDs sampled out to restframe ~160µm and SEDs that additionally include the sub-mm. They find that in the absence of submm data dust masses are a factor ≤ 2 more uncertain, but exhibit no net bias. Magdis et al. (2012b) confirm similar results for a small sample of high redshift galaxies detected in the sub-mm, excluding any data point at wavelength >200µm. Based on a Monte Carlo analysis, the errors on the stacked photometry correspond to a median uncertainty of 0.14dex for our dust masses. Berta et al. (in prep.) present a more comprehensive analysis of uncertainties in Herschel-based dust masses. In total, we obtain Draine & Li (2007) dust masses and Magnelli et al. (2014) dust temperatures for 512 bins in z, SFR and M*.

As recognized by several authors, there can be systematic differences between Draine & Li (2007)-based dust masses and the typically smaller ones derived using single-temperature modified black body models, with details depending on the treatment of dust opacities and emissivity indices (e.g., Magnelli et al. 2012a, Magdis et al. 2012b, Bianchi 2013, Berta et al. in prep.), as well as between dust temperatures that are based on different conventions. We defer a detailed discussion to Berta et al. (in prep) but note that Eq. (9) below and our subsequent results refer to dust masses from the Draine & Li (2007) method, dust temperatures as in Magnelli et al. (2014), and star formation rates based on the Wuyts et al. (2011a) ladder. For all bins, these SFRs agree within ±0.3dex with the IR luminosities from the stacks (Magnelli et al. 2014).

If dust is a calorimeter of the incident UV, radiating at an average temperature and optically thin in the far-infrared, a simple scaling is expected between $M_{dust}$, $T_{dust}$, and SFR. For our adopted scales and an emissivity index β=1.5, this relation takes the form

$$\left(\frac{M_{dust}}{M_\odot}\right) = 1.2 \times 10^{15} \left(\frac{SFR}{M_\odot yr^{-1}}\right) \times \left(\frac{T_{dust}(MBB)}{K}\right)^{-5.5} \qquad (9).$$

where the constant has been calibrated on the data for the 512 bins.

The conversion to gas masses requires the application of a metallicity dependent dust to gas ratio correction, which also enters the redshift evolution through the redshift dependence of the mass-metallicity relation below (e.g., Bethermin et al. 2014). Following Magdis et al. (2012b) and Magnelli et al. (2012a) we converted the Draine & Li (2007) model dust masses to (molecular) gas masses by applying the metallicity dependent dust to gas ratio fitting function for z~0 SFGs found by Leroy et al. (2011),

$$\delta_{dg} = \frac{M_{dust}}{M_{molgas}} = 10^{(-2+0.85\times(12+\log(O/H)-8.67))} \qquad (10),$$

where $12 + \log(O/H)$ again is the gas phase oxygen abundance (see also Draine et al 2007 for dust-to-gas with metallicity scalings of the SINGS nearby galaxy sample, and Galametz et al (2011) or Rémy-Ruyer et al. (2014) for lower metallicity galaxies down to 12+log(O/H)=8.0). We note that the metallicity dependence in equation (10) is within a few percent of that found in the last section from averaging equations (6) and (7). This means that the metallicity (and hence, mass) corrections we choose in this paper for the dust and CO data are very similar.

As in the case of the CO sample, the 512 stacks are grouped into 6 redshift bins comparable to those of the CO sample. These 512 stacks provide a complete and unbiased estimates of the mean FIR/submm properties of all SFGs with 0.16<z<2, $M_*$> $10^{10}$ $M_\odot$, and *log(sSFR/sSFR(ms,z,M*))*>-0.3 (see Figs. 4 & 5 of Magnelli et al. 2014). In contrast to the CO sample, the dust sample has comparable numbers of SFGs (83-191) in the middle 4 redshift bins, while the number in the lowest and highest bins are significantly smaller (~30 each), introducing substantially greater uncertainties in the redshift scaling relation for the dust data as compared to the CO data. This difference actually turns out to be advantageous in the discussion of the scaling relations below, as the dust sample is obviously not dominated in number by the lowest redshift bin.

### 2.3 Mass-metallicity relation

For the few SFGs in this paper with estimates of gas phase metallicities from strong line rest-frame optical line ratios, we determine individual estimates of $\log Z = 12 + \log (O/H)$. For instance, if the $\lambda 6583$ [NII]/ $\lambda 6563$ H$\alpha$ line flux ratio is measured the Pettini & Pagel (2004) indicator yields

$$12 + \log(O/H)_{PP04} = 8.9 + 0.57 \times \log(F(6583\,[NII])/F(6563\,H\alpha)) \qquad (11).$$

The scatter in the above relation is ±0.18 dex. However, for the large majority of the SFGs in our CO and dust samples, such line ratios are not available and it is necessary, for the metallicity corrections discussed above, to refer to the mass-metallicity relation. Following Maiolino et al. (2008) we combined the mass-metallicity relations at different redshifts presented by Erb et al. (2006), Maiolino et al. (2008), Zahid et al. (2014) and Wuyts et al. (2014) in the following fitting function

$$12 + \log(O/H)_{PP04} = a - 0.087 \times (\log M_* - b)^2, \text{ with}$$
$$a = 8.74(0.06), \text{ and}$$
$$b = 10.4(0.05) + 4.46\,(0.3) \times \log(1+z) - 1.78(0.4) \times (\log(1+z))^2 \qquad (12).$$

Mannucci et al. (2010) have presented evidence for a dependence of metallicity on star formation rate for z~0 SDSS galaxies, at a given stellar mass (the 'fundamental metallicity relation'), yielding an alternative version of equation (12),

$$\Delta Z_{M08} = 12 + \log(O/H)_{M08} - 8.69 = 0.21 + 0.39 \times x - 0.2 \times x^2 - 0.077 \times x^3 + 0.064 \times x^4,$$
$$\text{with } x = \log M_* - 0.32 \times \log SFR - 10, \text{ and}$$
$$12 + \log(O/H)_{PP04} - 8.9 = -0.4408 + 0.7044 \times \Delta Z_{M08} - 0.1602 \times (\Delta Z_{M08})^2$$
$$-0.4105 \times (\Delta Z_{M08})^3 - 0.1898 \times (\Delta Z_{M08})^4 \qquad (12a),$$

where stellar masses are in solar masses and star formation rates in solar masses per year. The Mannucci et al. relation (their equation 4) is on the Maiolino et al. (2008, M08) scale, which is then converted to the PP04 scale using the coefficients in their Table 4.

This relation implies that at constant stellar mass and within ±0.6 dex of the main-sequence the PP04 metallicity changes by -+0.04dex near solar metallicity, which is a second order correction for equations (6) and (7), given the uncertainties in metallicity and star formation rates (even for SFGs far from the main sequence the correction is only ~-0.1dex). At high-z several recent studies of 'strong emission line indicators' also indicate a weak dependence of metallicity on star formation rate for massive ($\log(M_*/M_\odot)>10$) galaxies (Steidel et al. 2014, Wuyts et al. 2014, Sanders et al. 2014). Wuyts et al. (2014) find that the fundamental metallicity relation in equation (12b) does broadly trace the redshift evolution of the mass-metallicity relation in equation (12). At constant z Wuyts et al. find that for a 0.6dex change in star formation rate the implied metallicity does not change more than 0.08dex, at $<z>=0.9$ and 2.3. While the application of the strong emission line indicators to metallicity determinations at high-z is currently under debate (e.g. Steidel et al. 2014), these results at face value imply a negligible correction in equations (6) and (7) when applying equation (12b). For these reasons our default assumption in the following is equation (12). We will discuss in section 4.2 how replacing equation (12) by (12a) quantitatively affects the scaling relations.

## 2.4 Stellar masses and star formation rates

For the COLDGASS sample the stellar masses and luminosities are calibrated in the frame of SDSS, GALEX and WISE (Saintonge et al. 2011a, 2012). For the various LIRG/ULIRG samples at z~0-1 we refer to the original papers for a discussion of the stellar masses, which were converted, if necessary, to the Chabrier IMF adopted here. Infrared luminosities were obtained from the far-infrared (30-300μm) SEDs, assuming $L_{IR}=1.3 \times L_{FIR}$ and star formation rates were estimated from Kennicutt (1998) with a correction to the Chabrier IMF adopted here, SFR ($M_\odot$ yr$^{-1}$) $=10^{-10} \times L_{IR}$ ($L_\odot$) (see discussion in Genzel et al. 2010).

At z≥0.1 global stellar properties for all optically/UV-selected SFGs (both for the CO and dust-samples) were derived following the "ladder of indicators" procedure as outlined by Wuyts et al. (2011a). In brief, stellar masses were obtained from fitting the rest-UV to near-IR spectral energy distributions (SEDs) with Bruzual & Charlot (2003) population synthesis models, the Calzetti et al. (2000) reddening law, a solar metallicity, and a range of star formation histories (in particular including constant SFR, as well as exponentially declining or increasing SFRs with varying e-folding timescales). SFRs were obtained from rest-UV+IR luminosities through the Herschel-Spitzer-calibrated ladder of SFR indicators of Wuyts et al. (2011a) or, if not available, from the UV-optical SED fits. For the main-sequence population (with near constant star formation histories) we adopt uncertainties of ±0.15 dex for the stellar masses, and ±0.2 dex for the star formation rates, although somewhat smaller uncertainties may be appropriate for SFGs with measurements of individual far-infared luminosities (Wuyts et al. 2011a).

For submillimeter galaxies (SMGs) we adopted the stellar masses and luminosities of Magnelli et al. (2012b, 2014, priv.comm.), the latter being derived from PACS/SPIRE Herschel SEDs and converted to star formation rates with the modified Kennicutt (1998) conversion as given above. The stellar masses and star formation rates of most above main-sequence outliers (ULIRGs, SMGs) are more uncertain than those of the main-sequence populations (±0.3 dex). The outliers are more dusty (e.g. Wuyts et al. 2011b), making population synthesis analysis of the UV/optical rest-frame SEDs more uncertain. The bursty nature of the star formation histories adds substantial additional uncertainties to the average star formation rates (e.g. Figure 6 in Genzel et al. 2010), as well as to the inferred stellar masses, as is demonstrated by the up to one order of magnitude varying estimates of stellar masses in SMGs in different publications (Hainline et al. 2009, 2011, Michalowski et al. 2012, 2014, Davè et al. 2010, Bussmann et al. 2012). In the specific case of the stellar masses for the GOALS LIRG/ULIRG sample used in this paper (Howell et al. 2010), this may result in an overestimate of the intrinsic stellar masses. AGN may contribute to the bolometric luminosity for the extreme starburst population (e.g. for z~0 ULIRGs, Genzel et al. 1998); this means that star formation rates in these systems may be overestimated (by 0.1..0.4 dex, see Genzel et al. 2010 and references therein).

Note that throughout the paper we define stellar mass as the "observed" mass ("live" stars plus remnants), after mass loss from stars. This is about 0.15…0.2 dex smaller than the integral of the SFR over time.

# 3. Results

Several recent papers have attempted to quantify the dependence of galaxy integrated molecular gas depletion time scale (or its inverse, often called the "star formation efficiency"), and the related molecular gas fraction, on redshift, specific star formation rate and stellar mass. For instance, from COLDGASS and PHIBSS 1 CO data Tacconi et al. (2013) infer the logarithmic scaling index with redshift, $\xi_{f1} = d(logt_{depl})/d(log(1+z))$, to range between -0.7 and -1, while Santini et al. (2014) find $\xi_{f1} \sim -1.5$ from PEP/Hermes Herschel dust data. Magdis et al. (2012b), Saintonge et al. (2012), Tacconi et al. (2013), Sargent et al. (2014) and Huang & Kauffmann (2014) all find that at a given redshift the depletion time scale decreases with increasing specific star formation rate relative to its value at the main-sequence line. The corresponding logarithmic scaling index, $\xi_{g1}=d(logt_{depl})/d(log(sSFR))$, ranges between -0.3 and -0.5. However, the exact value of $\xi_{g1}$ is strongly degenerate with variations of the CO conversion factor with *sSFR* (Magdis et al. 2012b). From a re-analysis of the COLDGASS sample Huang & Kauffmann (2014) find that the depletion time scale depends little on stellar mass or stellar mass surface density, once the dependency on specific star formation rate is removed.

The following analysis, based on a combination of similar size, CO and dust samples covering a comparable range in redshift, specific star formation rate and stellar mass, promises to permit a major step forward in delineating these principal component dependences and, in particular, the role of the CO conversion factor/function.

## 3.1 Separation of variables

In the left panel of Figure 2 we plot for the 6 redshift bins (different colored symbols) the CO-based depletion time scale as a function of normalized specific star formation rate offset from the main sequence line at a given redshift (*sSFR(ms,z,M*)*, equation (1)). In this log-log presentation $log(t_{depl})$ appears to scale **linearly** with $log(sSFR/sSFR(ms))$ over more than 3 orders of magnitude in *sSFR*, from more than a factor of 10 below, to two orders of magnitude above the main-sequence, and even in the regime of the extreme outliers, such as z~0-0.5 ULIRGs and some SMGs. This means that the dependence of $t_{depl}$ on *sSFR* is fit by a single power law, to within the uncertainties dictated by the scatter of the relation. We note that the conclusion of a single power law is strictly correct only if $\alpha_{CO}$ does not vary significantly with $log(sSFR/sSFR(ms))$, as has been assumed implicitly in equation (8). We explore the justification of this assumption for the near-main-sequence SFGs quantitatively in section 4.1 below. For the extreme above main-sequence, outlier/starburst ULIRG population at z~0 ($log(sSFR/sSFR(ms))>1$) there is persuasive evidence from mass modeling that $\alpha_{CO}$ is 4 to 5 times lower (Scoville, Yun & Bryant 1997, Downes & Solomon 1998, Daddi et al. 2010b, Genzel et al. 2010). If this correction is applied to the data in Figure 2, the resulting $log(t_{depl})$ - $log(sSFR/sSFR(ms))$ distribution would show a downward kink and would no longer be fit by a single power-law. This may imply a second 'starburst' mode of star formation (Sargent et al. 2012, 2014, Magdis et al. 2012b).

Fitting a power law to each of the $t_{depl} - (sSFR/sSFR(ms,z,M*))$ z-bins shows no significant redshift evolution of the slope $\xi_{g1}(z) = d(log\ t_{depl}(z) / dlog\ (sSFR/sSFR(ms,z,M*))$ (Figure 3). A weighted fit to the CO slope data in Figure 3 (filled blue circles) yields $d\xi_{g1}(z)/dlog(1+z)) = -0.08\ (\pm 0.13,\ 1\sigma)$. We note already here that there is also no evidence for a z-dependence of the dust-based depletion time scales (black filled circles in Figure 3) discussed below (section 3.2), for which a weighted fit to the 6 redshift bins yields $d\xi_{g1}(z)/dlog(1+z) = +0.13\ (\pm 0.18,\ 1\sigma)$.

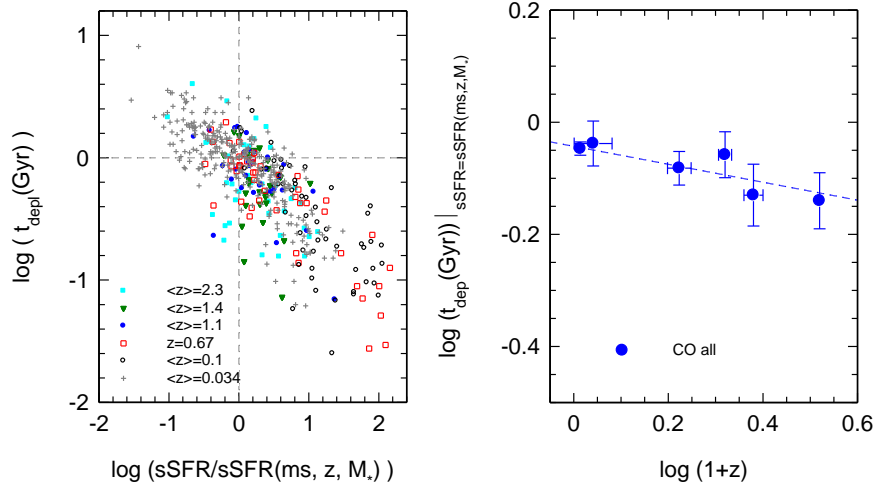

Figure 2: left panel: dependence of the CO-based molecular gas depletion time scale, $t_{dep}=\alpha_{0J} L_{CO}'/SFR$ (equations (4) & (8)) as a function of specific star formation rate, normalized to the main-sequence mid-line value at each redshift (from equation (1), Whitaker et al. 2012), for the 500 galaxies from Figure 1 with integrated CO measurements, binned in 6 redshift ranges from z=0 to 2.3. Right panel: dependence of the depletion time at the main-sequence mid-line on redshift, obtained from the zero-point offsets in slope -0.46 linear fits in the log-log distributions in the left panel in each redshift bin. The best linear fit has a slope of -0.16 (dashed line).

A similar analysis in the $log\ t_{depl} - log\ M_*$ shows that a linear function (a power law in the original variables) with slope 0 ($\pm 0.1$) can account for the data in each of the redshift bins.

These are important constraints. If a function of three independent variables (z, $sSFR/sSFR(ms,z,M*)$, $M_*$) is a power law in each of these variables, each with slope that does not depend on the other variable, then the function can be written as a ***product of power law functions each dependent only on one variable.*** That means that the variables can be separated,

$$t_{depl}(z, sSFR, M_*)|_{\alpha=\alpha_{0j}} = f_1(z)|_{sSFR=sSFR(ms,z,M_*)} \times g_1(sSFR / sSFR(ms, z, M_*)) \times h_1(M_*) \quad (13).$$

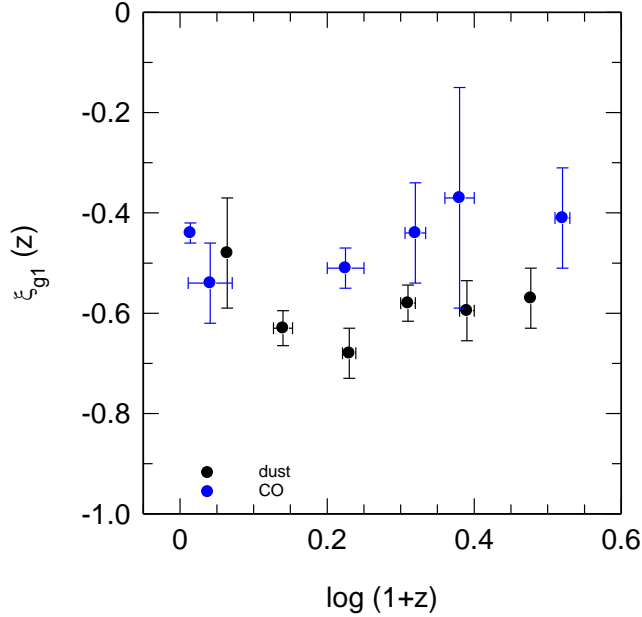

Figure 3. Slopes $\xi_{g1}(z) = dlog\, t_{depl}(z) / dlog\, (sSFR/sSFR(ms,z,M_*))$ as a function of log (1+z) for the CO data in the 6 redshift bins at $<z>\sim$0, 0.1, 0.67, 1, 1.4 and 2.3 (Table 1) (filled blue circles), and for the Herschel dust data in the 6 redshift bins at $<z>\sim$0.16, 0.35, 0.65, 1, 1.45 and 2 (Table 2) (filled black circles). Error bars are 1σ. A weighted power law fit to these data yields a slope of $d\xi_{g1}(z)/d\,log(1+z)= -0.08$ (±0.13, 1σ) for the CO data, and $d\xi_{g1}(z)/d\,log(1+z)= +0.13$ (±0.18) for the dust data. The best fitting constant (z) slopes $\xi_{g1}$ are -0.46 and -0.59 for the CO and dust data.

Here $f_1(z)$ tracks the dependence of $t_{depl}$ on redshift at the main-sequence line (equation (1)), $g_1(sSFR/sSFR(ms,z,M_*))$ describes the dependence of $t_{depl}$ on sSFR relative to the main-sequence line, and $h_1(M_*)$ delineates the stellar mass dependence. Now at first glance, the way we have written equation (13) (and analyzed the data) might seem a contradiction of the statement on variable separation above, since $g_1(sSFR/sSFR(ms,z,M_*)$ contains equation (1) in the denominator, which is a function of both z and $M_*$. However, equation (1) again is a product of power laws, so that the dependence of $sSFR(ms,z,M_*)$ can be easily pulled out of $g_1$. Equation (13) can then be resorted into a product of power laws of the individual variables (z, sSFR, $M_*$), as needed for separation. The slope of $g_1$ remains unaffected by this renormalization. It turns out that because of the shallow mass dependence of equation (13), the mass dependence also does not change much when resorting $g_1$ as a function of sSFR only. The only function strongly affected is $f_1(z)$ since that now acquires a strong redshift dependence from equation (1). This discussion already suggests that the separation of variables and the general conclusions on the quality of fits are independent of the choice of the main-sequence prescription, which we will discuss in more detail in section 4.2. The depletion time scale may also depend on other parameters, such as bulge mass, gas volume and surface density, environmental density around the galaxy etc. but these dependencies cannot be explored with the current data sets.

If the parameter dependencies of $t_{depl}$ can be separated according to equation (13), equation (3) shows that the parameter dependencies of molecular gas fractions can be separated as well,

$$\frac{M_{molgas}}{M_*}(z, sSFR, M_*)|_{\alpha=\alpha_{0J}} = t_{depl} \times sSFR$$

$$= f_2(z)|_{sSFR=sSFR(ms,z)} \times g_2(sSFR/sSFR(ms, z, M_*)) \times h_2(M_*) \quad (14).$$

We thus assume
$\log(g_1(sSFR/sSFR(ms, z, M_*))) = a_{g1} + \xi_{g1} \times \log(sSFR/sSFR(ms, z, M_*))$ (c.f. Saintonge et al. 2011b, 2012). In the first iteration we took the best fit constant slope from Figure 3 ($\xi_{g1}$=-0.46) and fitted the data in each redshift bin for the zero offset $a_{g1}(z)$. The resulting z-distribution of the zero offsets is shown in the right panel of Figure 2. Their z-dependence can again be well described by a power law, $\log(f_1(z)) = a_{f1} + \xi_{f1} \times \log(1+z)$.

In the second iteration we then removed the fitted zero-point offsets as a function of redshift by dividing the original data by $f_1(z)$ and fit the specific star formation rate dependences with a power law function, as above, but now for all 500 data points simultaneously. This has the obvious advantage of giving a much more robust estimate of the slope $\xi_{g1}$. It also strengthens our assumption that the data set can be fit by a single power-law. The result is shown in the right panel of Figure 2 and the left panel of Figure 4. We find that the best fitting parameters are $a_{f1}$=-0.043 (±0.01), $\xi_{f1}$=-0.16 (±0.04), $a_{g1}$=0, $\xi_{g1}$=-0.46 (±0.03). The numbers in the parentheses are the statistical 1σ fit uncertainties only; they do not include uncertainties due to systematics and cross-terms in the co-variance matrix.

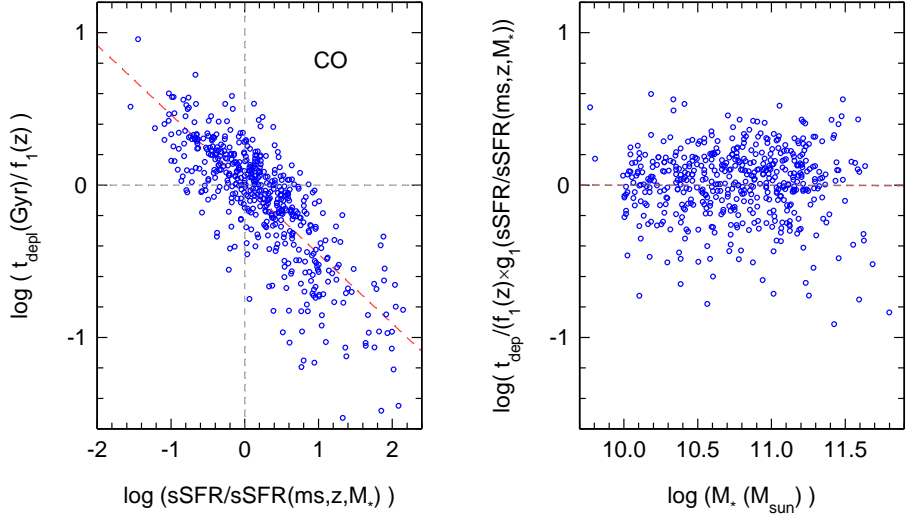

Figure 4: left panel: dependence of CO-based depletion time scale (equations (4) & (8)) on specific star formation rate normalized to the mid-line of main-sequence at each redshift (equation (1) and Figure 3, Whitaker et al. 2012), after removing the redshift dependence with the fitting function $f_1(z)=10^{-0.04-0.16\times log(1+z)}$ obtained from the right panel of Figure 3. The red-dashed line is the best linear fit to the log-log distribution of all 500 SFGs and has a slope of -0.46. The residuals have a scatter of ±0.24 dex. Right panel: dependence of CO-based depletion time scale on stellar mass, after removing also the specific star formation rate dependence with the fitting function $g_1(sSFR/sSFR(ms,z,M_*))=10^{-0.46\times log(sSFR/sSFR(ms,z,M_*))}$ obtained from the left panel of the Figure.

The redshift dependence is shallower than found by Tacconi et al. (2013, $\xi_{f1}$=-0.7… -1). This is entirely explainable by the fact that we now quote the redshift dependence of the depletion time *on* the main-sequence line. Tacconi et al. (2013) instead took an average of the COLDGASS data ($<t_{depl}>_{COLDGASS}$ ($z$=0)=1.5 Gyr) and the $z$=1.2 PHIBSS data ( $<t_{depl}>_{PHIBSS}$ ($z$=1.2) =0.6-0.7 Gyr). In that case the logarithmic slope is -1. The COLDGASS sample has many massive galaxies below, while PHIBSS1 has mostly galaxies above the main sequence line, such that the average depletion times are biased high (at $z$=0) and low (at $z$=1.2) because of the negative slope in the left inset of Figure 2, resulting in an overestimate of the redshift dependence. The slope in the *log $t_{depl}$ – log sSFR/sSFR(ms,z,M*)* plane inferred above is in agreement with Saintonge et al. (2012) and Huang & Kauffmann (2014) from COLDGASS data alone.

One might be concerned that the observed correlation between *sSFR/sSFR(ms,z,M*)* and $t_{depl}$ is caused (or at least affected) by the fact that specific star formation rate is proportional to, and the depletion time is inversely proportional to star formation rate, introducing an artificial negative correlation if the *SFR* has a substantial uncertainty. To explore the impact of this effect, we created Monte Carlo mock data sets. We started with the scaling relations in Table 3 ($t_{depl}$ as a function of $sSFR/sSFR(ms,z,M_*)$) for a mock data set spanning an order of magnitude in stellar mass and centered around some redshift, and then adding ±0.2 (respectively ±0.3 dex for outliers) scatter in *SFR*. We find that the artificial anti-correlation between *sSFR* and $t_{depl}$ is only significant if the data have near constant *sSFR*. For data with a range in *sSFR* comparable to our observed sample (2.5…3.5 dex), the effect merely leads to a very small increase in scatter but does not change the slope of the intrinsic relation.

The dispersion of the data in the left panel of Figure 4 around the best fitting power law function is ±0.24 dex. This is quite tight given the uncertainties of stellar masses (±0.15 to ±0.3 dex), star formation rates (±0.2 to ±0.3 dex), and molecular gas masses (>±0.2 dex), and considering the possibility of substantial variations of the CO conversion factor across the more than 3 orders of magnitude $sSFR/sSFR(ms,z,M_*)$ variation spanned by the data in Figure 4. There is a tendency for the $log(t_{depl}/f_1)$ residuals as a function of $log(sSFR/sSFR(ms,z,M_*))$ to exhibit an excess of negative values for $log(sSFR/sSFR(ms,z,M_*))>0.6$. This may indicate that the depletion time scale above the main sequence, in the starburst-outlier regime, drops faster than captured by the power law fit above.

The right panel of Figure 4 explores whether the residuals $log(t_{depl}/(f_1 \times g_1))$ depend on the remaining internal parameter, stellar mass. There appears to be no significant trend, in excellent agreement with the findings of Huang & Kauffmann (2014) at z~0.

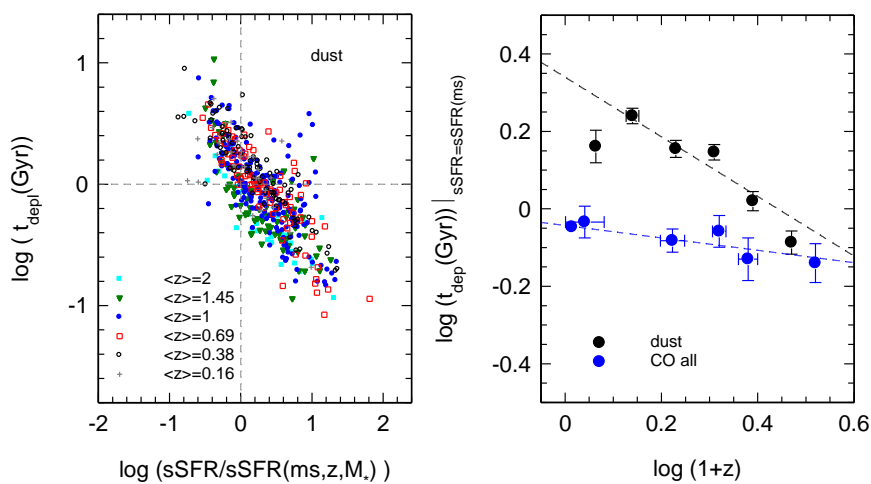

Figure 5: left panel: dependence of the dust-based molecular gas depletion time scale, $t_{dep}=M_{molgas}(M_{dust})/SFR$ (equations (9) & (10)) as a function of specific star formation rate, normalized to the main-sequence mid-line value at each redshift (from equation (1),

Whitaker et al. 2012), for the 512 PACS-SPIRE stacks from Magnelli et al. (2014), binned in 6 redshift ranges from z=0.16 to 2. Right panel: dependence of the dust-based depletion time (black circles) at the main-sequence mid-line on redshift, obtained from the zero-point offsets in slope -0.59 linear fits in the log-log distributions in the left panel in each redshift bin. The best linear fit has a slope of -0.77 (black dashed line). For comparison the filled blue circles and blue dashed line denote the CO-based data from Figure 3.

### 3.2 Dust based determination of $t_{depl}$

Next we repeat the same exercise for the dust based depletion time estimates from Herschel. The left panel of Figure 5 again shows the depletion time measurements as a function of *sSFR/sSFR(ms,z,M*)* in the 6 redshift bins. As for CO, the dust based depletion time scales do not exhibit a significant redshift evolution of $\xi_{g1}(z) = d(\log t_{depl}(z) / d\log (sSFR/sSFR(ms,z,M_*))$ (black filled circles in Figure 3), so that a constant slope $\xi_{g1}$=-0.59 is an adequate description of the current dust data. The dependences of $t_{depl}$ on redshift, specific star formation rate and stellar mass can again be written as a product of power laws, as in equations (13) and (14). Proceeding as before, we determine the zero-points of these power law functions in each bin, and plot these zero points as a function of redshift in the right panel of Figure 5 (black filled circles), along with the zero points of the CO-based determinations from Figure 2 (blue filled circles).

These totally independent estimates are *remarkably close*, especially given the possibly hidden systematic uncertainties, in CO conversion factor on the one hand, and in dust modeling and conversion from dust to gas on the other. The dust-based depletion time appears to vary faster with redshift ($\xi_{f1}$=-0.77 (±0.19)) than the CO-data ($\xi_{f1}$=-0.16 (±0.04)). The difference is only moderately significant (~3σ), since the dust sample in the lowest and highest redshift bins each only contains two dozen data points and since the *z*-coverage is smaller than in the CO-data (0.16< *z*<2). The extrapolated (*z*=0) zero point of the dust data ($a_{f1}$=0.34 (±0.07)) is larger than that of the CO data ($a_{f1}$=-0.04 (±0.01)). However, on average between z=0 and 2.5 the CO- and dust-based depletion time estimates do not differ by more than ~30%. We note that an even better and more straightforward comparison would be the direct comparison of gas masses determined in the same galaxies from each of the two techniques. This route is not possible, however, since the two samples do not significantly overlap, and in addition the dust technique involves stacking a number of galaxies, rather than observing individual galaxies.

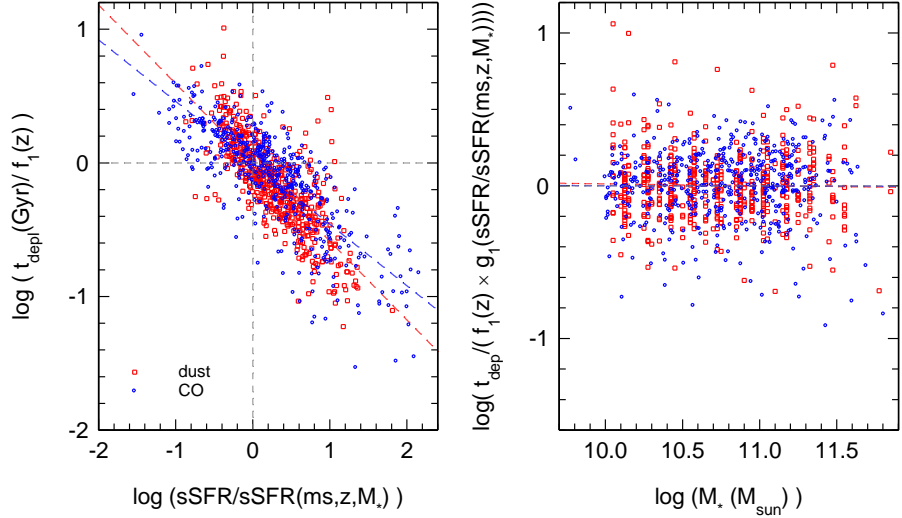

Figure 6: left panel: dependence of CO-based (blue) and dust-based (red) depletion times on specific star formation rate normalized to the mid-line of main-sequence at each redshift(equation (1), Whitaker et al. 2012), after removing the redshift dependences with the fitting functions $f_1(z)$ given in the right panels of Figure 3 and Figure 5. The blue- and red-dashed lines are the best linear fits to the log-log distributions of all 500 CO SFGs and all 512 Herschel stacks. Right panel: dependence of CO-based (blue) and dust-based (red) depletion times on stellar mass, after removing also the specific star formation rate dependence with the fitting function
$g_1(sSFR/sSFR(ms,z,M_*)) = 10^{-0.46 \times log(sSFR/sSFR(ms,z,M_*))}$ for CO and
$g_1(sSFR/sSFR(ms,z,M_*)) = 10^{-0.59 \times log(sSFR/sSFR(ms,z,M_*))}$ for the dust data, as obtained from the left panel of the Figure. The residuals from the best power law fits are ±0.24 dex for both the CO- and dust-data.

The similarity between CO- and dust-based scaling relations is further strengthened when comparing the redshift-corrected dependencies of depletion time scale on specific star formation rate offset, for CO- (blue circles) and dust-data (red squares) in the left and right panels of Figure 6. The slope of the dust data in the specific-star formation rate scaling relation is somewhat steeper than that of the CO-data (dust: $\xi_{g1}$ =-0.59 (±0.05), CO: $\xi_{g1}$=-0.46 (±0.03)) but the overall distributions overlap over almost three orders of magnitude in $sSFR/sSFR(ms,z,M_*)$, from below the main sequence to the starburst outliers. The lack of a trend as a function of stellar mass also agrees in dust and gas (right panel of Figure 6). This excellent agreement of the separated scaling relations is also particularly relevant because of the very different redshift distributions of the CO- and dust-data in terms of numbers of SFGs per redshift bin. While the CO data in Figure 6 are heavily weighted toward the z~0 COLDGASS measurements, the dust data are strongly weighted to z~1. The agreement in scalings with *sSFR* and $M_*$ thus cannot be an artefact of biased redshift distributions.

Table 3 summarizes the fit parameters for the power law fitting functions for the CO- and dust-data. It also gives the parameters for an average between the gas and dust relations that might be considered the best current description of the molecular gas depletion time scaling relations.

### *3.2.1. Global fits and error estimates*

For an alternative evaluation of the fits and their errors we went back and re-analyzed the CO- and dust-data in a different way. Instead of first binning in z-bins we carried out a direct global fit to data, assuming that the depletion times could be modelled as a linear function in the 3-space *log(1+z)*, *log(sSFR/sSFR(ms,z,$M_*$))* and *log$M_*$*. We then repeatedly fitted the 500 CO- and the 512 dust-data points, in each iteration perturbing the *sSFR/sSFR(ms,z,$M_*$)* and $t_{depl}$ values by ±0.2 dex for the main-sequence galaxies, and ±0.3 dex for the above main-sequence outliers (section 2.4). The fit results are in excellent agreement with the method discussed in section 3.1, indicating that the fit results are robust and the quoted errors are well captured by the error of the underlying parameters. We list these global fits values as the second row in each of the entries of Table 3.

## 3.3 Scaling relations for $M_{molgas}/M_*$

Next we determined the equivalent relations for the molecular gas fractions. As discussed in the Introduction, once the scaling relations for $t_{depl}$ are established, those for $M_{molgas}/M_*$ in principle follow straightforwardly from equations (1) and (3). Given the slopes of the best fitting power laws in Table 3 one would then expect for the equivalent power law $M_{molgas}/M_*$-fitting function from equations (1), (3) and (14),

$$\log(M_{molgas}/M_*(z, sSFR, M_*)|_{\alpha=\alpha_{0J}}) =$$
$$= \log(f_2(z)|_{sSFR=sSFR(ms,z,M_*)}) + \log(g_2(sSFR/sSFR(ms,z,M_*))) + \log(h_2(M_*))$$
$$= a_{f2} + \xi_{f2} \times \log(1+z) + \xi_{g2} \times \log(sSFR/sSFR(ms,z,M_*)) + \xi_{h2} \times \log(M_*) \quad (15),$$

with slopes $\xi_{f2} = \xi_{f1} + 3.2 \sim +2.9$, $\xi_{g2} = \xi_{g1} + 1 \sim +0.5$, and $\xi_{h2} = \xi_{h1} - 0.3 \sim -0.3$.

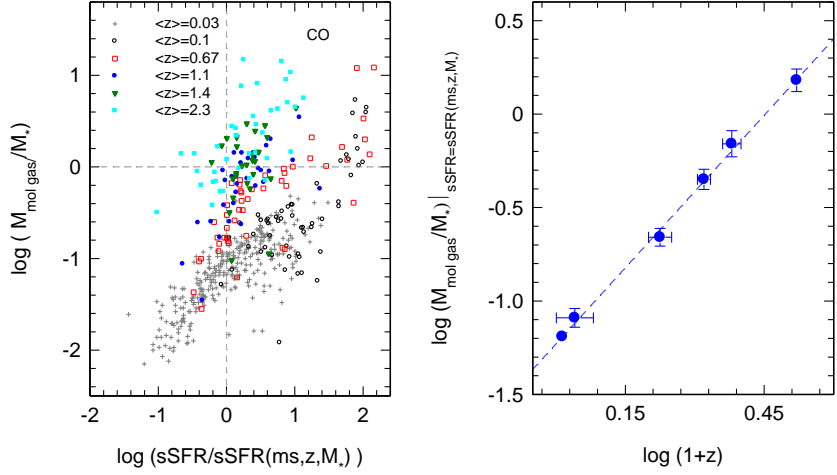

Figure 7. left panel: dependence of the CO-based molecular gas mass to stellar mass ratio, $M_{molgas}/M_* = \alpha_{0\,J} L_{CO\,J}'/M_*$ (equations (4) & (8)) as a function of specific star formation rate, normalized to the main-sequence mid-line value at each redshift (from equation (1), Whitaker et al. 2012), for the 500 galaxies from Figure 1 with integrated CO measurements, binned in 6 redshift ranges from z=0 to 3.4. Right panel: dependence of the CO-based molecular gas mass to stellar mass ratio at the main-sequence mid-line on redshift, obtained from the zero-point offsets in slope +0.51 linear fits in the log-log distributions in the left panel in each redshift bin. The best linear fit has a slope of 2.71 (dashed line).

In order to check for systematic effects, we decided to check the consistency of the scaling relations (together with equation (1)) for the CO and dust samples, by computing $M_{molgas}/M_*$ for each galaxy (or galaxy stack) and then establish the scaling relations. Figures 7 to 10 show this fitting carried out for the CO data (Figures 7 & 8) and for the dust data (Figures 9 & 10).

Consistent with the results of the last section, the agreement between the CO- and dust-data again is very good. Again we fitted the data also with the global fit method described in section 3.2.1., establishing that both the best fit values and their uncertainties are robust and well captured by the most probable errors (statistical plus systematic) of the underlying parameters. The parameters of the best fitting power law functions are summarized in Table 4. We find $\xi_{f2}$=2.7, $\xi_{g2}$=0.5 and $\xi_{h2}$=-0.4 for both the "binned" and the global fitting methods.

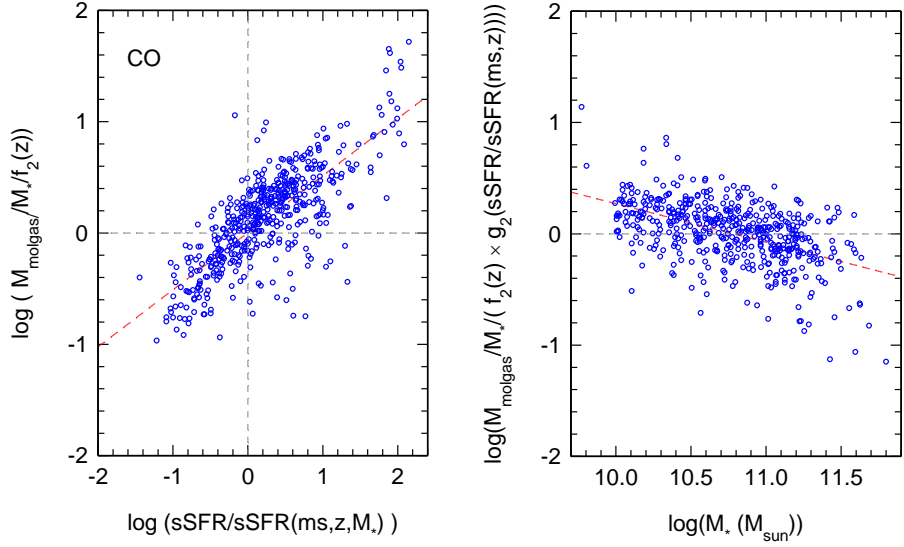

Figure 8: left panel: dependence of CO-based molecular gas mass to stellar mass ratio, $M_{molgas}/M_* = \alpha_{0\,J} L_{CO\,J}'/M_*$ (equations (4) & (8)) on specific star formation rate normalized to the mid-line of main-sequence at each redshift (equation (1), Whitaker et al. 2012), after removing the steep redshift dependence with the fitting function $f_2(z)=10^{-1.23+2.71 \times log(1+z)}$ obtained from the right panel of Figure 7. The red-dashed line is the best linear fit to the log-log distribution of all 500 SFGs and has a slope of 0.51. Right panel: dependence of CO-based gas mass to stellar mass ratio on stellar mass, after removing also the specific star formation rate dependence with the fitting function $g_2(sSFR/sSFR(ms,z,M_*))=10^{0.51 \times log(sSFR/sSFR(ms,z,M_*))}$ obtained from the left panel of the Figure. The resulting best linear fit has a slope of -0.35.

These slopes (and also the zero points) are within ~0.1 dex of the expectations from equation (3), and give a measure of the internal systematic uncertainties. We will come back to this topic in section 4.2.

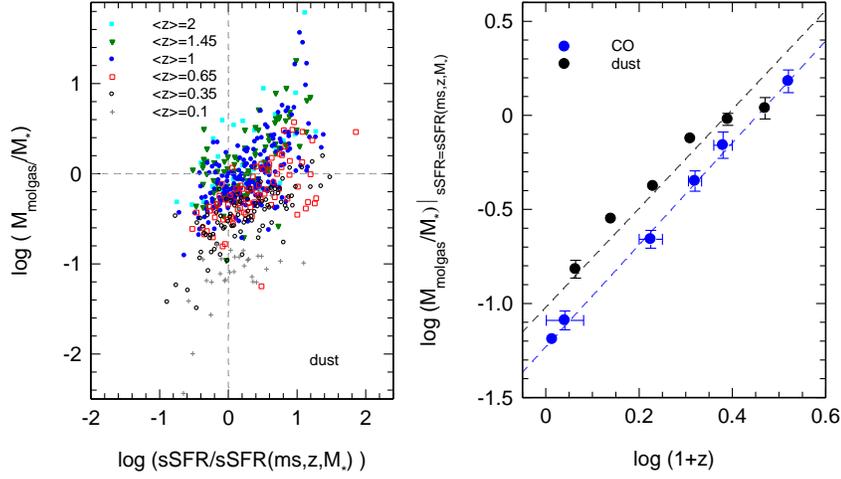

Figure 9: left panel: dependence of the dust-based molecular gas mass to stellar mass ratio, $M_{molgas}/M_* = M_{molgas}(M_{dust})/M_*$ (equations (9) & (10)) as a function of specific star formation rate, normalized to the main-sequence mid-line value at each redshift (from equation (1), Whitaker et al. 2012), for the 512 PACS-SPIRE stacks from Magnelli et al. (2014), binned in 6 redshift ranges from z=0.1 to 2. Right panel: dependence of the dust-based molecular gas mass to stellar mass ratio (black circles) at the main-sequence mid-line on redshift, obtained from the zero-point offsets in slope 0.5 linear fits in the log-log distributions in the left panel in each redshift bin. The best linear fit has a slope of 2.26 (black dashed line). For comparison the filled blue circles and blue dashed line denote the CO-based data from Figure 8.

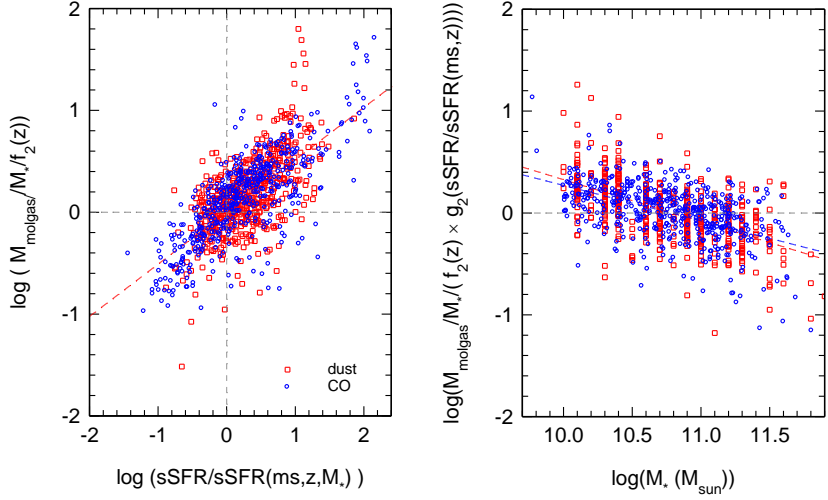

Figure 10: left panel: dependence of CO-based (blue) and dust-based (red) molecular gas mass to stellar mass ratios on specific star formation rate normalized to the mid-line of main-sequence at each redshift (equation (1), Whitaker et al. 2012), after removing the redshift dependences with the fitting functions $f_2(z)$ given in the right panels of Figure 7 and Figure 9. The blue- and red-dashed lines are the best linear fits to the log-log distributions of all 500 CO SFGs and all 512 Herschel stacks. Right panel: dependence of CO-based (blue) and dust-based (red) molecular gas to stellar mass ratio, after removing also the specific star formation rate dependence with the fitting function $g_2(sSFR/sSFR(ms,z,M_*))=10^{-0.51 \times log(sSFR/sSFR(ms,z,M_*))}$ for CO and for the dust data, as obtained from the left panel of the Figure.

## 3.4 Intermediate Summary

The depletion times and molecular gas to stellar mass ratios derived from two independent and very different techniques (CO and dust), with each ~500 measurements covering the redshift range from 0 to 2-3, the range in specific star formation rate from below to much above the main-sequence line at each redshift, and in stellar mass from $10^{10}$ to several $10^{11}$ $M_\odot$ yield similar zero points and, to within the uncertainties, the same scaling indices. We note here again that a more direct cross-check of the two techniques through a direct, galaxy-by-galaxy comparison would be highly desirable, but is currently not possible because of the lack of sample overlap. The good agreement is better than we would have expected (but see Magdis et al. 2012b). Given the multiple parameter dependences of the CO- and dust-techniques, one might easily have predicted offsets and trends between these techniques of 0.2-0.5 dex. We have only corrected the CO mass estimates for excitation and metallicity dependences, and the dust to gas mass ratios for metallicity dependence. Given the systematic parameter dependences and uncertainties of the calibrations used in each of the two techniques the good agreement (in zero point and logarithmic scaling indices) thus is gratifying.

On this basis, our analysis yields the following main results

- to first order, the dependences on redshift, specific star formation rate and stellar mass can be well separated into a product of three power law functions depending on the individual parameters;
- the depletion time on the main sequence line is smaller than the Hubble time at all z, and changes only slowly with redshift ($t_{depl} \propto (1+z)^{-0.3\pm0.15}$, by a factor of 0.7 from z=0 to z=2.5), in broad agreement with our earlier study in Tacconi et al. (2010, 2013) and less rapidly than found by Santini et al. (2014). The molecular gas to stellar mass ratios and the specific star formation rates as a function of redshift track each other fairly closely ($\propto (1+z)^3$), again in broad agreement with Tacconi et al. (2013), Magdis et al. (2012b), Saintonge et al. (2013), Santini et al. (2014) and Sargent et al. (2014). This finding in turn suggests that the factor of ~20-30 increase in galactic star formation rates between the local Universe and the peak of cosmic star formation rate at z~1-3 is mainly driven by the increased supply rate of fresh gas, rather than changes in galaxy scale star formation efficiency (in starbursts with small $t_{depl}$). This is consistent with the "gas-regulator" model (Bouché et al. 2010, Tacconi et al. 2010, 2013, Daddi et al. 2010, Davé et al. 2012, Lilly et al. 2013);
- changes in specific star formation rate at constant z and $M_*$ are due to a combination of variations in gas fraction and depletion time scale throughout the redshift range probed, in agreement with the z~0 COLDGASS results of Saintonge et al. (2012), Magdis et al. (2012b) and Huang & Kauffmann (2014). Galaxies above the main sequence have larger gas fractions but also at the same time smaller depletion time scales (or equivalently, higher star formation efficiency), in approximately equal measure, than galaxies at or below the main sequence. The dependence on gas fraction may reflect the time variation in the average gas supply rate from the cosmic web. The increase in 'star formation efficiency' with *sSFR* (by a factor of 20 from the lower envelope of the main-sequence to the star-bursting outliers above the main-sequence) may be driven by the internal properties of the star forming interstellar medium, such as the dense gas fraction (Gao & Solomon 2004, Lada et al. 2012, Juneau et al. 2009, Gracia-Carpio et al. 2011, Elbaz et al. 2011) in the more compressed, cuspier SFGs above the main sequence (Wuyts et al. 2011b, Elbaz et al. 2011). The increasing depletion times below the main sequence, especially at high masses, may also be a signature of suppression of the gravitational instability by large shear velocities driving up the Q of the ISM ("morphological quenching": Martig et al. 2009, Genzel et al. 2014, Tacchella et al. in preparation).
- throughout the redshift range probed the molecular gas to stellar mass ratios decrease as a function of stellar mass ($M_{gas}/M_* \propto M_*^{-0.4}$) but the depletion time scale does not vary with stellar mass, in agreement with Tacconi et al.

(2013), Santini et al. (2014), Huang & Kauffmann (2014) and Sargent et al. (2014). This is in contrast to the ideal gas regular model ($d(sSFR)/d(logM_*)|_{main\ sequence} \sim 0$, Lilly et al. 2013), for which no or only little dependence of gas fractions on stellar mass would be expected. This means that the dropping gas fractions of the observed SFGs at and above the Schechter mass ($log(M_S/M_\odot) \sim 10.9$) are a direct consequence of the fact that the observed SFGs on the main sequence have lower specific star formation rates than expected for an ideal gas regulator ($sSFR$=const). We interpret these findings as empirical evidence for the expected quenching process(es) that are theoretically expected to happen near and above the Schechter mass.

# 4. Discussion

## 4.1 Parameter scalings of the CO-molecular gas mass conversion factor

Our results so far are based on the tenet that the ratio of molecular gas mass to CO luminosity needs only a correction for metallicity and ladder excitation (equation (8)); ***$α_{CO}$ is assumed to not vary with z, sSFR and $M_*$.*** This approach was motivated in section 2.1.1. We will now return to the issue of variations of the CO conversion factor/function as a function of these parameters.

In this section we will take advantage of the fact that the dust data are not dependent on a conversion factor, although they obviously do depend on other uncertain prescriptions for the metallicity dependence of the gas to dust ratio, as well as for the mass-metallicity relation and the Draine & Li (2007) dust emissivity modeling. As such, we can plausibly assume that the dust data can be treated as 'ground truth' for the impact of $α_{CO}$ variations.

A sensitive and straightforward, robust test of the dependence of $α_{CO}$ on $sSFR, z$ and $M_*$ can be derived from the well determined dependence of dust temperature on $sSFR/sSFR(ms,z,M_*)$ in the Herschel dust data of Magnelli et al. (2014) (see also Magdis et al. 2012b, Santini et al. 2014). Of course, such a test can also be made from the comparison of the dependence of depletion time estimates as a function of $sSFR/sSFR(ms,z,M_*)$ in Figure 6 (with very similar results) but the test on the dust temperature distribution is more robust, ***as the latter does not depend implicitly on the specific dust model (i.e. Draine & Li 2007), or the prescriptions for the dust to gas ratio and stellar mass as function of metallicity***. As we will see below, the dust to gas ratio prescription appears explicitly in the equations, and the dust modeling enters only in the zero point of the dust mass estimates (the constant in equation 9).

Following Magnelli et al. (2014) and Magdis et al. (2012b) the relation between molecular gas depletion time scale and dust temperature in the limit of optically thin far-IR dust emission (and optically thick dust absorption in the UV/optical, dust as a 'calorimeter') is given by (see equation 9)

$$\frac{L_{IR}}{M_{dust}} = c_1 \times \frac{SFR}{M_{molgas} \times \delta_{dg}(Z)} = c_1 \times \frac{1}{t_{depl} \times \delta_{dg}(Z)} = c_2 \times T_{dust}^{4+\beta} \qquad (16),$$

where the metallicity dependent dust to molecular gas ratio $\delta_{dg}$ is given in equation (10), $\beta$ is the logarithmic scaling index of the frequency dependence of dust emissivity ($\beta \sim 1.5$, Magnelli et al. 2014), and $c_1$, $c_2$ (as well as $c_3$, $c_4$, $c_5$ below) are constants. We now expand the CO-conversion function (for J=1) as a function of the four main parameters to the linear terms in the log,

$$\log(\alpha_{CO\,1}) = \log(\alpha_{MW}) + \log(\alpha_{CO\,1\,0}/\alpha_{MW})$$
$$+ \varepsilon_z \times \log(1+z) + \varepsilon_Z \times (\log(Z) - \log(Z_\odot))$$
$$+ \varepsilon_{sSFR} \times \log(sSFR/sSFR(ms,z,M_*))$$
$$+ \varepsilon_* \times (\log M_* - 10.5) \qquad (17),$$

where $\varepsilon_z, \varepsilon_Z, \varepsilon_{sSFR},$ and $\varepsilon_*$ are the logarithmic slopes of the variations of $\alpha_{CO}$ as a function of redshift, metallicity, specific star formation rate relative to that at the main sequence, and stellar mass offset from $\log M_* = 10.5$. We define $\alpha_{CO\,1\,0}$ as the 'zero point' of the conversion factor, that is, its value at $z=0$, $sSFR = sSFR(ms,z,M_*)$, $\log M_* = 10.5$ and $Z = Z_\odot$. The metallicity dependence is assumed to follow equations (6) and (7) with $\varepsilon_Z \sim -1.2$. Next we can write

$$\log t_{depl} = \log\left(\frac{\alpha_{CO} \times L'_{CO}}{SFR}\right) = \log t_{depl}|_{\alpha_{CO\,1} = \alpha_{MW}} + \log(\alpha_{CO\,1\,0}/\alpha_{MW})$$
$$+ \varepsilon_z \times \log(1+z) + \varepsilon_Z \times (\log(Z) - \log(Z_\odot))$$
$$+ \varepsilon_{sSFR} \times \log(sSFR/sSFR(ms,z,M_*))$$
$$+ \varepsilon_* \times (\log M_* - 10.5) \qquad (18),$$

where the left side of the equation is the 'true' depletion time scale as given in equation (16), while the first term on the right side is the 'observed' depletion time scale under the assumption that $\alpha_{CO} = \alpha_{MW} =$ const, as in Table 3 and equation (13),

$$\log t_{depl}|_{\alpha_{CO\,1} = \alpha_{MW}} = a_{f1} + \xi_{f1} \times \log(1+z) + \xi_{g1} \times \log(sSFR/sSFR(ms,z,M_*)) +$$
$$\xi_{h1} \times (\log(M_*) - 10.5) \qquad (19).$$

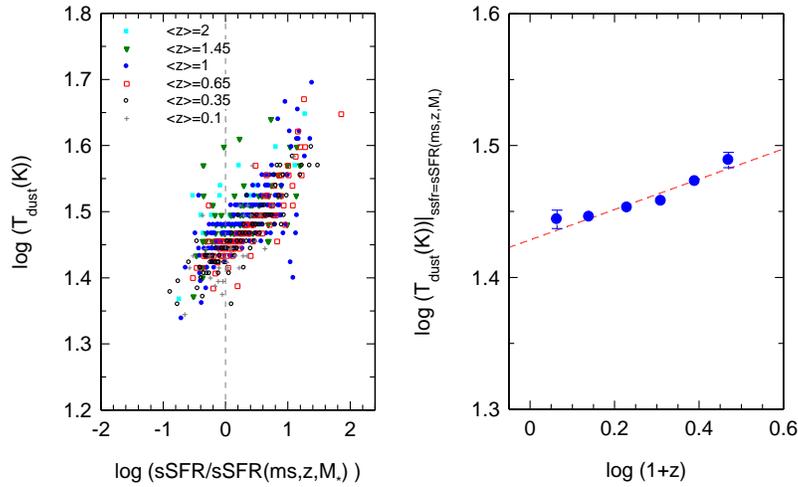

Figure 11: left panel: dependence of the dust temperature as a function of specific star formation rate, normalized to the main-sequence mid-line value at each redshift (from equation (1), Whitaker et al. 2012), for the 512 PACS-SPIRE stacks from Magnelli et al. (2014), binned in 6 redshift ranges from z=0.1 to 2. Right panel: dependence of the dust temperature at the main-sequence mid-line on redshift, obtained from the zero-point offsets in slope 0.086 linear fits in the log-log distributions in the left panel in each redshift bin. The best linear fit has a slope of 0.11 (black dashed line).

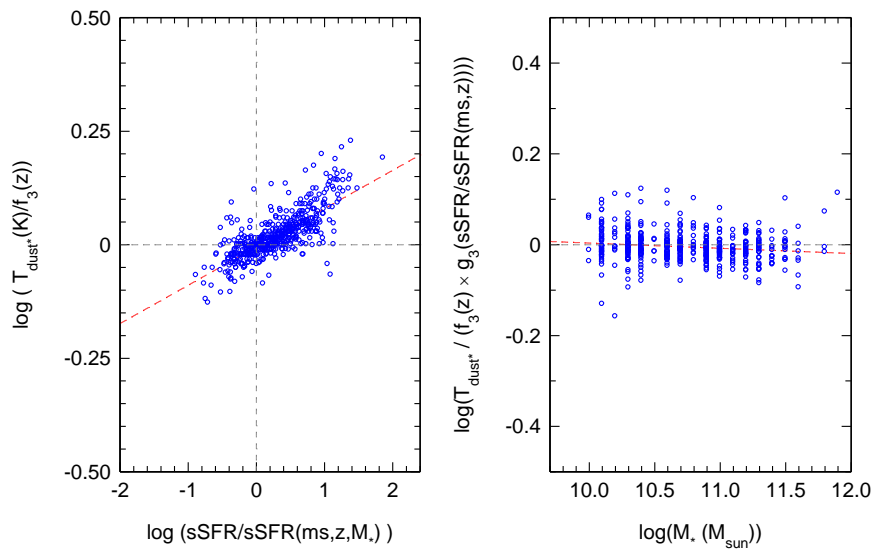

Figure 12: left panel: dependence of dust temperature on specific star formation rate normalized to the mid-line of main-sequence at each redshift (equation (1), Whitaker et al. 2012), after removing the redshift dependence with the fitting function $f_3(z)=10^{+1.43+0.11\times\log(1+z)}$ obtained from the right panel of Figure 7. The red-dashed line is the best linear fit to the log-log distribution of all 512 stacks and has a slope of 0.086 (±0.003). Right panel: dependence of dust temperature on stellar mass, after removing also the specific star formation rate dependence with the fitting function $g_3(sSFR/sSFR(ms,z,M_*))=10^{0.064\times\log(sSFR/sSFR(ms,z,M_*))}$ obtained from the left panel of the Figure. The resulting best linear fit has a slope of -0.012.

Proceeding as before for depletion time scale and gas to stellar mass ratio, the dependence of the observed dust temperature on redshift, specific star formation rate offset and stellar mass can also be separated into the product of three power law functions (as for $t_{depl}$ and $M_{molgas}/M_*$) to yield

$$\log T_{dust} = a_{f3} + \xi_{f3} \times \log(1+z) + \xi_{g3} \times \log(sSFR/sSFR(ms,z,M_*)) + \xi_{h3} \times (\log(M_*)-10.5) \quad (20).$$

Figures 10 & 11 show the scaling relations of the dust temperature in the same 512 galaxy stacks as in Figures 5 & 6 and 9 & 10, and Table 5 summarizes the best fit values for the power law fitting function in equation (20).

Equations (16), (18), (19) and (20) can now be combined to yield

$$\log(T_{dust}) = c_3 - \frac{1}{4+\beta} \times \begin{bmatrix} (\xi_{dg}+\varepsilon_Z)\times(\log Z - \log Z_\odot) + (\xi_{f1}+\varepsilon_z)\times\log(1+z) + \\ (\xi_{g1}+\varepsilon_{sSFR})\times\log(sSFR/sSFR(ms,z,M_*)) + \\ (\xi_{h1}+\varepsilon_*)\times(\log M_* - 10.5) \end{bmatrix}$$

$$= c_4 + \xi_{f3}\times\log(1+z) + \xi_{g3}\times\log(sSFR/sSFR(ms,z,M_*))$$
$$+ \xi_{h3}\times(\log M_* - 10.5) \quad (21),$$

which after sorting finally results in

$$0 = (0.35(\pm 0.14) - \log(\alpha_{CO\,1\,0}/\alpha_{MW}) +$$
$$(-\xi_{dg}-\varepsilon_Z)\times(\log Z - 8.67) +$$
$$(-\xi_{f1}-(4+\beta)\times\xi_{f3}-\varepsilon_z)\times\log(1+z) +$$
$$(-\xi_{g1}-(4+\beta)\times\xi_{g3}-\varepsilon_{sSFR})\times\log(sSFR/sSFR(ms,z,M_*)) +$$
$$(-\xi_{h1}-(4+\beta)\times\xi_{h3}-\varepsilon_*)\times(\log M_* - 10.5) \quad (22).$$

This shows that the logarithmic slopes of the dependence of $\alpha_{CO}$ on $Z$, $z$, $sSFR/sSFR(ms,z,M_*)$ and $M_*$ can be uniquely constrained from the scaling relations of $t_{depl}$ and $T_{dust}$. If equation (22) is to be fulfilled everywhere in the sampled parameter space and the different variables on the right hand side are independent, each of the coefficients in front of the variables on the right hand side must be zero.

*Constraints for main-sequence population:* with the fit results in Tables 3 and 5, and the assumption that the Herschel dust observations provide "ground truth" we find for the near-main sequence population

$$\alpha_{CO\,1\,0} = 9.8\ (\pm 3.2)$$
$$\varepsilon_Z = -0.9\ (\pm 0.3)$$
$$\varepsilon_z = -0.4\ (\pm 0.13)$$
$$\varepsilon_{sSFR} = 0\ (\pm 0.03)$$
$$\varepsilon_* = 0.07\ (\pm 0.02) \quad (23).$$

The inferred metallicity dependence of the CO-conversion factor is broadly consistent with equations (6) & (7) ($\varepsilon_Z \sim -1.2 \pm 0.3$). The derived redshift dependence suggests that the conversion factor at z~2.2 is ~0.7 times that at z~0, which would of course then imply a steeper gradient ($\xi_{f1}$ ~ -0.5), in agreement with the dust depletion time evolution in Figure 5. The fact that the inferred zero point of the conversion factor is twice $\alpha_{MW}$ is also consistent with the z=0 shift between dust- and CO-depletion times in Figure 5. This last conclusion is strongly dependent on the dust modelling assumptions in Draine & Li (2007). For instance, for a simple modified black body modelling the dust masses would come down by a factor of ~0.5-0.7 dex (Magnelli et al. 2012a, Berta et al., in prep.), in that case resulting in a zero point of $\alpha_{CO\,1\,0}$ = 2 to 3.

In our opinion the most important result is that the CO conversion factor near the main sequence depends little on *sSFR*. This is consistent with the earlier studies of Magdis et al. (2012a) and Magnelli et al. (2012a) but with much improved statistical confidence. Across *Δlog(sSFR/sSFR(ms,z,M∗))* = ±0.6 the dust temperature measurements set an upper limit to a change in the CO conversion factor of 6% for the mean value, and of 25% if the 2σ uncertainties are included. This is a strong constraint, which applies as long as the dust measurements indeed provide a 'ground-truth' estimate.

*Constraints for above main sequence outliers*: the dust temperature gradient in Figure 12 appears to steepen ($\xi_{g3}$ increases from 0.086 to ~0.135) for the outliers above the main sequence (up-bend of the distribution of points in the left panel for *Δlog(sSFR/sSFR(ms,z,M∗))* ≥ +0.9), implying greater radiation field densities than on the main sequence (see Magdis et al. 2012b). At the same time the CO- and dust-depletion times in this regime are comparable (no significant change in $\xi_{g1}$ (CO) ~ -0.46), albeit with increased scatter (left panel of Figure 6). Equation (22) then implies a drop in the conversion factor ($\varepsilon_{sSFR} = -\xi_{g1} - (4+\beta) \times \xi_{g3}$). At *Δlog(sSFR/sSFR(ms,z,M∗))* ~ +1..+1.3 the residuals from the slope 0.086 power law fit (red-dotted line in Figure 12) are about +0.05 dex, which implies a decrease of the conversion factor to $\alpha_{CO}$~2 ($\varepsilon_{sSFR}$ ~ -0.3). Within the 0.3 dex uncertainties this is consistent with $\alpha_{CO}$~0.8-1.5, the value empirically estimated by Downes & Solomon (1998) and Scoville et al. (1997) for z~0 ULIRGs (see also Bolatto et al. 2013). If such a change of $\alpha_{CO}$ is applied, depletion time scales for the extreme outliers decrease by 0.3-0.7 dex, suggesting the presence of a second, 'starburst' mode, as discussed by Daddi et al. (2010b), Genzel et al. (2010), Magdis et al. (2012 a,b) and Sargent et al. (2014). The main quantitative difference to Magdis et al. (2012 a,b) and Sargent et al. (2014) is that our data perhaps suggest that these deviations in $\alpha_{CO}$ become significant only about one order of magnitude, and not, as in these papers, already a factor of 4 above the main-sequence line. The statistics of our data in Figures 6 and 12, however, are not sufficient to distinguish a continuous from an abrupt change in $\alpha_{CO}$, and one needs to keep in mind the large uncertainties of stellar masses and star formation rates for this obscured, bursty population (section 2.4).

Finally the last constraint on the stellar mass dependence implies a change of $\alpha_{CO}$ of less than 7% from $10^{10}$ to $10^{11}$ M$_\odot$ in stellar mass.

## 4.2 Discussion of uncertainties

### *4.2.1. Final global fits*

In order to create the best final estimates of the scaling relations, we finally averaged/combined the CO- and dust-based relations/data, which are listed under the title 'average' in the last section of Tables 3 and 4, now under the assumption that these data sets provide two independent estimates of 'ground truth'. This 'averaging' effectively means that we are using a solar metallicity 1-0 conversion factor of $\alpha_{MW} \times 10^{0.1}$=5.5 (for the Draine & Lee 2007 modeling). For the 'binned' method discussed in sections 3.1 and 3.2 we averaged the fit values of the scaling relation obtained with the two methods, respectively. For the 'global' fit column in Table 3 we first added 0.1 dex as a zero point correction to all CO-, and subtracted 0.1 dex from all dust-depletion time and $M_{gas}/M_*$ values, to bring the CO and dust data on a common zero point, before then carrying out a global fit to all 1012 data points, as described in section 3.2.1. Inspection of Tables 3 and 4 demonstrates that, to within the uncertainties of about ±0.24 dex, ***the fit results are quite robust to the changes in fitting technique and whether or not CO and dust data are used separately or combined.*** We recommend the global fits as our currently best estimates of the scaling relations (bold face numbers in Tables 3 & 4).

However, given these systematic uncertainties and the varying selection functions in the data used in our analysis, there are indeed differences at this level. This can be seen, for instance, by comparing the gas masses computed from the depletion time scaling relation in Table 3 to those computed from the gas to stellar mass ratio scaling relations in Table 4. On average the latter are about 25% (0.1 dex) larger than the former, and larger source to source variations are possible in different parts of the parameter space, owing to cross terms in the relations of Table 3 & 4 and equation (1). Likewise the redshift dependence of the gas mass to stellar mass ratio ($\xi_{f2}$= 2.7, Table 4) is less by 0.16 (±0.1) dex than what one would have expected from the redshift dependence of the depletion time scale ($\xi_{f1}$= -0.34), the redshift dependence of the specific star formation rate in Whitaker et al. (2012, equation (1)), $d(sSFR(ms,z,M_*))/d(log(1+z))$ ~3.2 and equation (3). Because of the lack of correlation of the depletion time scale with stellar mass, and its shallow dependence on redshift, we recommend using the fitting equations in Table 3, multiplied by *SFR*, for calculating gas masses.

### *4.2.2. Dependence of the results on the prescription of the main-sequence*

A significant source of uncertainty comes from the choice of the fitting function for the main-sequence *sSFR(ms,z,M$_*$)* (equation (1). As we have stated in section 1, the Whitaker et al. (2012) fitting function used throughout this paper does trace the location of the observed SFGs in this paper, as well as its parent samples between z=0 and 2.5 and at logM$_*$>10.2, quite well. However, it over-predicts *SFR* and *sSFR* at lower stellar masses (which are not sampled in this paper), where a more accurate prescription has been proposed by Whitaker et al. (2014). There are significant variations in the main-sequence prescriptions proposed in the literature, depending on galaxy selection criteria, star formation and mass tracers used etc. (Schiminovich et al. 2007, Noeske et al. 2007, Elbaz et al. 2007, 2011, Daddi et al. 2007, Panella et al. 2009, Rodighiero et al. 2010,

Peng et al. 2010, Karim et al. 2011, Salmi et al. 2012, Lilly et al. 2013). These variations introduce differences in the specific star formation rate of the main sequence line, as well as in particular in the stellar mass dependence of the main sequence at a given redshift.

We have explored what happens if instead of the Whitaker et al. (2012) fitting function, the simpler function proposed in equation (2) of Lilly et al. (2013) is chosen. That function is a shallow single power law as a function of stellar mass ($sSFR(ms,z,M_*) \sim M_*^{-0.1}$), without a redshift dependence of the slope (or curvature) as in the Whitaker et al. (2012, 2014) prescriptions. As a result it does somewhat better below $10^{10}$ M$_\odot$ but above $log(M_*/4 \times 10^{10} M_\odot)$ the Lilly et al. (2013) function predicts too high star formation rates and overshoots the observed locus of SFGs. The Whitaker and Lilly fitting functions bracket other prescriptions proposed in the literature. We repeated the global fitting with the Lilly et al. (2013) prescription and list the resulting fit parameters for the scaling laws in depletion time scale and gas to stellar mass ratio in the third to last columns of Tables 3 and 4. With the exception of modest changes in the overall zero points and in the logarithmic slopes as a function of stellar mass (expected because of the relative locations of SFGs and fit at high mass discussed in the last sentences), the differences to the fits with the Whitaker prescriptions are almost negligible. This is especially relevant for the dependence on the parameter $sSFR/sSFR(ms,z,M_*)$. This relative insensitivity to the main-sequence-prescription is because the values of $sSFR(z,M_*)$ in Whitaker et al. (2012, 2014) and Lilly et al. (2013) agree quite well between $M_* = 2$ to $10 \times 10^{10}$ M$_\odot$, where most of our SFGs reside. The differences increase outside this mass range, where we have few galaxies ($< 2 \times 10^{10}$ M$_\odot$), or where these differences then lead to a slightly different slope in the $logM_*$ scaling relation ($\xi_{h1}$ parameter in the last column of Tables 3 & 4).

### 4.2.3. Fitting in sSFR-space

Instead of using a main-sequence prescription, it is also possible to express $g_1$ and $g_2$ as function of $sSFR$ directly. These global fit results are listed in the second to last columns of Table 3 & 4. Since no main-sequence prescription is involved in this case $\xi_{f2} = \xi_{f1} \sim +1.2$, $\xi_{g2} = +0.5 = \xi_{g1} + 1$, and $\xi_{h2} = \xi_{h1} \sim -0.2$, as expected. The global fits in the third to last ($g_{1,2}(sSFR/sSFR(ms,z,M_*))$) and the second to last columns in Tables 3 and 4 ($g_{1,2}(sSFR)$) have the same formal scatter of ±0.24 dex, consistent with the expected systematic uncertainties. As such the fits are equivalent. The main difference is in the interpretation of the resulting depletion time scales as a function of redshift, at constant specific star formation rate and stellar mass. In the $sSFR$-description the depletion time scale appears to be a strong function of redshift ($\xi_{f1} \sim +1.2$). At constant $sSFR$ galaxy at z=2 has a 3.8 time greater depletion time scale as at z=0. This is not a physical effect, however, since specific star formation rates vary strongly with redshift. For instance a $SFR$=100 M$_\odot$ yr$^{-1}$ "ULIRG" SFG at z=0 is an extreme outlier starburst above the main sequence, with a correspondingly short depletion time scale, while an SFG of the same SFR at z=2 is a common main sequence object near equilibrium.

### *4.2.4. Impact of metallicity descriptions*

A different choice in the metallicity dependences of the CO conversion factor (equations (6) & (7)) and of the dust to gas ratio (equation (10)) will also have a significant impact of the final parameter values. These metallicity corrections were calibrated at z=0 and are very uncertain below about 0.3 solar metallicity, which affects stellar masses below about $10^{10}$ $M_\odot$, especially at z>1. The Leroy et al. (2011) calibration of the dust to gas ratios is strictly applicable only to z=0 and refers to the total (molecular plus atomic) gas column; its redshift evolution is not known. However, with the exception of a few SFGs at z>2 the stellar mass range of our sample implies a modest metallicity range (~0.2 dex for 97% of our SFGs) from slightly below to slightly above solar, resulting in a range of CO conversion factors of less than a factor of 2. The same is true for the dust-to-gas ratios. Because of this fairly limited metallicity (or mass) range sampled in the galaxies considered here, a different metallicity scaling might change the zero point but the impact on scaling relations should be second order.

We have checked what changes occur if the redshift dependent, mass-metallicity relation for estimating metallicities in equation (12) is replaced by the fundamental metallicity relation of Mannucci et al. (2010), which depends on both stellar mass and star formation rate (equation (12a)). That equation does not have an explicit redshift dependence, but implicitly the z-dependence comes in through the star formation rates. The effect of using equation (12a) instead of (12) are on average slightly lower metallicities (by about 0.07 dex), and thus modest upward corrections of the gas masses inferred by either the CO- and the dust-technique. As a result the zero point of the depletion time increases by 17% relative to our default global model. When comparing equation (12a) to (12) the differential correction factor decreases with redshift. As a result the redshift evolution for the depletion time scale is slightly steeper for the FMR. Other than that the differences between these metallicity prescriptions have no significant impact on the scaling relations in Tables 3 and 4.

### *4.2.4 Systematic uncertainties and missing parameters*

It is important to recall that both CO- and dust- methods rely on several uncertain assumptions (cloud counting, near-virialized clouds, dust emissivity modelling) and have substantial systematic uncertainties (dust model, metallicity dependent corrections calibrated at z~0, mass-metallicity relation). The (rest frame far-infrared) dust observations, as well as the CO 2-1 and 3-2 data used for most of the high-z galaxies are only sensitive to T~20-40 K dust/gas in star forming regions. They do not trace cold (<10 K) gas and dust, or the atomic interstellar medium. The latter results in an important correction to the total galaxy gas contents at z~0 (~0.3-0.4 dex, Saintonge et al. 2011a,b) but may not play a dominant role at high-z (Lagos et al. 2012). The proper way to interpret the scaling relations discussed in this paper then is that they refer to ***the star forming gas***, and not to 'sterile' gas components in the outer parts of the galaxies, or in the galaxy disks but not participating in gravitational collapse.

Judging from the existing CO-ladder observations at both low and high z (see references in section 2.1.1.), the excitation corrections we have applied should be on average valid to 0.1 dex, at least for SFGs near the main sequence and for J≤3. The corrections are more uncertain for compact extreme starbursts (with highly excited ISMs), or for extended low temperature disk galaxies (with a significant component of <20 K gas that would be missed in the CO 3-2 transition).

If the errors of the input parameters are estimated correctly (±0.2 dex for each *sSFR/sSFR(ms,z,M\*)* and $t_{depl}$ for the main-sequence populations, and ±0.3 dex for the starburst outliers, all these considerations suggest that the 'average' relations in Tables 3 and 4 should give the scalings of the depletion time scale and molecular gas to stellar mass ratio between z=0 to 2.5, *log(sSFR/sSFR(ms,z,M\*))* between -1 and +1, and $log(M_*/M_\odot)$ from 10 to 11.5 to about ±0.1 dex in relative terms, and ±0.2 dex including systematic uncertainties.

### 4.3 Interpretation of the shallow redshift dependence of the depletion time

From a theoretical perspective, the slow change of the molecular gas depletion time with cosmic epoch ($t_{depl} \propto (1+z)^{-0.34\pm015}$) is somewhat surprising (e.g. Kauffmann, White & Guiderdoni 1993, Cole et al. 1994, Elmegreen 1997, Silk 1997). Considering again the definition of the depletion time in the context of the KS-relation (equation (2)), $t_{depl}$ might naturally be thought to be proportional to the galaxy's dynamical time, with the proportionality being the inverse of the galaxy's star formation efficiency $\eta$ (Kauffmann et al. 1993, Cole et al. 1994, Silk 1997, Elmegreen 1997, Kennicutt 1998, Genzel et al. 2010). In the Mo, Mao & White (1998) framework of disk formation in a dark matter dominated Universe, the disk's dynamical time $t_{dyn}(R_d)$ (expressed in terms of its scale length $R_d$ and maximum rotation velocity $v_d$) is tied to the properties of the dark matter halo, such that

$$t_{depl} = \frac{t_{dyn}(R_d)}{\eta} = \frac{1}{\eta} \times \frac{R_d}{v_d} = \frac{\lambda R_h}{\eta \times C_h v_h} = \frac{\lambda}{\eta C_h} \times 0.1 H(z)^{-1} \qquad (24).$$

Here $R_h$ and $v_h$ are the halo size and circular velocity, and $C_h$ is the ratio of the disk's rotation velocity to the halo circular velocity, which depends on the halo concentration (Bullock et al. 2001a). The parameter $\lambda$ is the angular momentum parameter of the baryons ($<\lambda>_{dark\ matter}$ ~0.04, Bullock et al. 2001b), and $H(z)=H_0 \times (\Omega_\Lambda + \Omega_0 \times (1+z)^3)^{1/2}$ is the Hubble parameter. In a matter dominated Universe (applicable at high-z) the depletion time should then be proportional to $(1+z)^{-3/2}$ (see Davé et al. 2011), which is inconsistent with our results. A more careful evaluation requires two corrections. First the Hubble parameter in a ΛCDM Universe changes more slowly at late times. If one approximates $H(z) \propto (1+z)^\beta$, the average β for the redshift range z=0 to 2.5 is -0.98. Second the concentration parameter of dark matter halos was smaller at high-z than at z=0 (Bullock et al. 2001a), such that $C_h$ ~ 1.025 at z~2.5 and 1.24 at z~0 (see Somerville et al. 2008). Taken together these two effects change the effective redshift dependence in equation (16) to $(1+z)^\beta$ with *β= -0.83* (again between z=0 and 2.5). In a recent evaluation of star forming disk sizes in CANDELS/3D HST between z=0 and 3 van der Wel et al. (2014) find empirically $\beta$= -0.75 (±0.05), which is smaller than $\beta$= -0.83 but comes close to it. Additional baryonic processes connected to the processing/dissipation of the angular momentum from the scale of the cosmic web to the inner, star forming disk and feedback processes inside the disk can both increase and decrease $\lambda$ of the baryonic component (Dutton et al. 2011, Danovich et al. 2014). Empirically the average observed $\lambda$ of the star forming gas at z~0.8-2.5 is about 0.035, which is (fortuitously) close to the dark matter $\lambda$ (Burkert et al. in preparation).

The difference between the values of van der Wel's -0.75 (±0.05) and our average estimate of -0.34 (±0.15) for the slope in the redshift dependence of the depletion time is not highly significant. If it were, the shallow slope may suggest that the depletion time scale is not set by the global galactic dynamical time but by local processes. There are good reasons for this view. Cloud collapse and star formation are local processes that proceed on the free fall time $\tau_{ff}$, which depends on the inverse square root of the local gas density (e.g. Krumholz & McKee 2005, Krumholz, Dekel & McKee 2012). Empirically Leroy et al. (2013) find from spatially resolved observations of the KS-relation in 30 z~0 disk galaxies that the depletion time is near constant and independent of the local or global orbital times. However, the high-z disk galaxies in our sample appear to be globally marginally stable systems with a Toomre parameter $Q \leq Q_{crit}$~1. In this case it is easy to show that the depletion time scale in the KS-relation on large scales is locked to the average galactic orbital time, even if in principal the local volumetric star formation rate density is tied to the local free fall time scale (e.g. Genzel et al. 2010, Krumholz, Dekel & McKee 2012). The dependence of depletion time scale on specific star formation rate ($t_{depl} \propto (sSFR/sSFR(ms,z,M_*))^{-1/2}$) may be considered as another argument in favour of a local origin of the depletion time scale.

### 4.4 Impact on the gas – star formation relation

The scaling relations in Tables 3 & 4 can be used to predict the form of the galaxy integrated molecular gas – star formation relations at different redshifts and under varying selection functions. Because of the strong dependence of the depletion time on specific star formation rate ($t_{depl} \propto (sSFR/sSFR(ms,z,M_*))^{-0.5}$) the relation between $M_{mol\,gas}$ and *SFR* (the "molecular" KS-relation, Kennicutt 1998) becomes super-linear for a sample of SFGs with a spread in *sSFR*, although intrinsically the relation at constant *sSFR* is linear.

We explored the impact of this dependence more quantitatively, by creating mock galaxy samples at different redshifts and with varying spread in $sSFR/sSFR(ms,z,M_*)$ and establishing the resulting $M_{mol\,gas}$-*SFR* from the relations in equation (1) and Table 3. For this purpose we repeatedly drew samples of 10-100 SFGs at a given z (the upper value is an upper limit to the samples available now and in the near future, at least for z>0.5). At each redshift we varied redshifts by ±10-30% around the mean redshift and varied stellar masses from $log(M_*/M_\odot)$=10.3 to 11.3. We also varied *SFR* and $M_{mol\,gas}$ estimates by ±0.2 dex to reflect measurement uncertainties, and varied $\Delta(log(sSFR/sSFR(ms,z,M_*))$ to sample the vertical direction in the $M_*$-SFR plane. The final mock sample gives a approximately uniform coverage in all these parameters.

At all z, the resulting integrated KS-slope N=*dlog(SFR)/d(M$_{mol\,gas}$)* in these mock data sets approaches ~1 for large N and a small range in *sSFR*. However, even for a pure main-sequence sample *Δ(log(sSFR/sSFR(ms,z,M$_*$))*=±0.3 the slope becomes super linear, N~1.1-1.2, and with a substantial scatter of ±0.07 dex resulting from the variation in SFG parameters. The scatter naturally increases with the decreasing number $n_G$ of galaxies in the sample, from ±0.06 for $n_G$ =50-100, to ±0.13 for $n_G$=20, to ±0.2 for $n_G$=10. The slope increases with the range of *sSFR*, from N=1.08 for *Δ(log(sSFR/sSFR(ms,z,M$_*$))*=±0.3, to N=1.33 for *Δ(log(sSFR/sSFR(ms,z,M$_*$))*=±0.6, and to N=1.6 for *Δ(log(sSFR/sSFR(ms,z,M$_*$))*=±1. Finally the slope for fixed *Δ(log(sSFR/sSFR(ms,z,M$_*$))* increases by about 0.2 from z=0 to z=2.

These findings are in excellent agreement with the integrated galaxy KS-slopes in the literature, both at low-z (Kennicutt 1998, Bigiel et al. 2008, Leroy et al. 2013, Saintonge et al. 2012, Sargent et al. 2014), as well at z≥1 (Daddi et al. 2010b, Genzel et al. 2010, Tacconi et al. 2013, Sargent et al. 2014). In particular the dependence of the slope on *Δ(log(sSFR/sSFR(ms,z,M$_*$))* is in excellent agreement with the findings of Kennicutt (1998), Daddi et al. (2010b), Genzel et al. (2010), Magdis et al. (2012), Sargent et al. (2014) and Tacconi et al. (2013). Note that for galaxy samples with extreme selection functions, such as a combination of a near-main sequence sample with a sample of starburst outliers (such as studied by Daddi et al. 2010b and Genzel et al. 2010), the continuous change in $t_{depl}$ with *sSFR* would then appear like the superposition of two separate gas-*SFR* relations, as proposed by Sargent et al.(2014).

For lack of size information we cannot study here the more common KS-surface density relation $\Sigma_{mol\,gas}$-$\Sigma_{SFR}$. Whether or not extra factors come into play in the surface density relation as compared to the integrated quantities depends on the underlying physical reason of the *sSFR*-dependence of $t_{depl}$ and the probably connected inverse scaling of sizes with *sSFR* (Wuyts et al. 2011b, Elbaz et al. 2011). Krumholz, Dekel & McKee (2012) have pointed out that if the KS-relation intrinsically is a volumetric relation between the gas and star formation volume densities, $\rho_{SFR} \propto \rho_{gas}/\tau_{ff}(\rho_{gas})$, the scale height $h_{gas}$ of the gas comes in as an additional factor, such that the galaxy averaged surface density relation becomes

$$\Sigma_{SFR} = <\rho_{SFR} \times h_{gas}> \propto h_{gas}^{-1/2} \times <\rho_{gas} \times h_{gas}>^{3/2} \propto h_{gas}^{-1/2} \times \Sigma_{gas}^{3/2} \qquad (25),$$

since the local free fall time scale $\tau_{ff}(\rho_{gas}) \propto \rho_{gas}^{-1/2}$. If the scaling $t_{depl} \propto sSFR^{-1/2}$ is physically connected with average local gas densities increasing with *sSFR*, as indicated by the findings of Wuyts et al. (2011b) and Elbaz et al. (2011), equation (25) would be in qualitative agreement with our scaling relation, with the addition of the scale height factor.

To summarize these results qualitatively: A intrinsically linear K-S relation at constant vertical position in the sSFR-$M_*$ plane transforms into a super-linear relation once there is a range of sSFRs, either in sample of galaxy integrated measurements, or in a spatially resolved sample within a galaxy for which the initial K-S relation is applicable. This is because at greater sSFR (or SFR, or $\Sigma_{SFR}$) the gas fraction increases (pushing a data point to the right in the K-S plane) AND the depletion time decreases (pushing the data point up in the K-S plane), resulting in a combined shift to the upper right along the diagonal. When data points with a range of sSFR (or SFR or $\Sigma_{SFR}$) are combined this appears as a super-linear relation with scatter.

## 4.5 Extrapolating to the future: dust or CO method?

We have shown in this paper that the derivation of molecular gas masses in star forming galaxies either from low-J CO rotational line emission (using a CO conversion 'function' that is dependent on metallicity, and contains a simple excitation correction), or from full far-infrared SEDs (with a conversion to gas mass that uses the Draine & Li (2007) dust model, a metallicity dependent gas to dust ratio from Leroy et al. (2011), and the mass-metallicity relation), yield consistent results across a wide range of *z*, *sSFR/sSFR(ms,z,M*)*, and *M*$_*$. We have argued that the combined scaling relations in Tables 3 to 5 capture the most important variations of molecular gas depletion time scale, molecular to stellar mass ratios and average dust temperature, within the ±0.24 dex, ±0.24 dex and ±0.033 dex scatter of the three relations, and to an absolute level of ±25%.

Extrapolating to future work in the measurement of cold gas masses in galaxies, it is likely that the focus will be on expansion of the statistics and parameter range on the one hand, including studies of the dependence on parameters that we have not been able to explore (such as galaxy morphology and environment). On the other hand spatially resolved measurements are likely to play an increasingly important role, especially at higher redshift, in order to explore the dependence of depletion time scale and gas fraction on internal galaxy structure, such as bulge to disk mass ratio, molecular volume and surface densities, clumpiness and internal galaxy kinematics, including galactic turbulence. The latter measurements will call for high resolution data benefitting naturally from or even requiring the availability of molecular gas kinematics that comes for free in line measurements of CO, HCN etc.

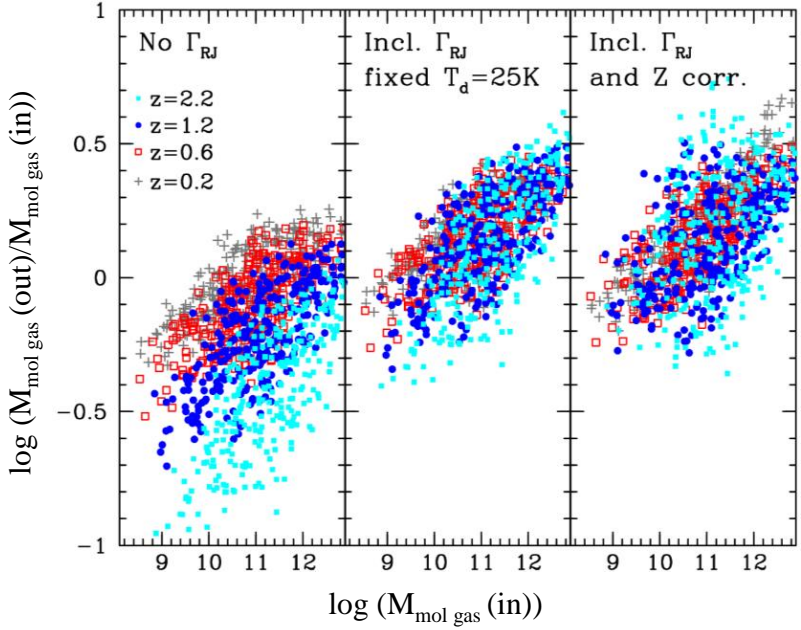

Figure 13: Gas masses inferred from single frequency photometry (ALMA band 7, 350 GHz) in the Rayleigh Jeans-tail of the dust SED (for assumed $T_{dust}$=25 K= const: the 'Rayleigh-Jeans tail' method, see text), relative to the "true" input gas masses. For this purpose we used the scaling relations in Tables 4 to 5 to compute input gas masses as a function of redshift, specific star formation rate offset and stellar mass ($\log(M_*/M_\odot)$=10-11.5), on the same grid points as in Magnelli et al. (2014, same color coding as in Figures 5, 9 and 11). The left panel shows the performance of the 'Rayleigh-Jeans tail' method (see Scoville 2013, Scoville et al. 2014) for $T_{dust}$=const in the Rayleigh-Jeans approximation (instead of applying a Planck correction with a dust temperature that is varying according to the scaling relations in Table 5), and without the metallicity dependent dust to gas ratio correction in equation (10). The central panel shows the performance of the 'Rayleigh-Jeans tail' method if a constant Planck correction (for $T_{dust}$=const=25 K) is applied to all data but again (as in the left panel) the metallicity dependent gas to dust ratio correction is omitted. The right panel still uses a constant Planck correction but now the metallicity dependent dust to gas ratio correction in equation (10) is applied.

The former goal primarily requires *efficient* measurements of galaxy integrated gas masses. To this end, it is well known that the detection of a given gas mass from broad-band detection of its submillimeter dust emission is substantially faster than from CO 2-1 or 3-2 data. In the case of ALMA the detection of a molecular gas mass of $10^{10}$ M$_\odot$ from band 7 (350 GHz) broad-band data requires only a few minutes at z~1-2 (for a 5σ detection), while a CO-based measurement (again at 5σ) requires more than one hour at z~0.6-1 and several hours at z~2 (Scoville 2013, Scoville et al. 2014)[3].

---

[3] The integration times quoted here and shown in Figures 14 & 15 are for 34 active 12 m antennas, 7.5 GHz bandwidth in dual polarization for the dust measurements and do not include overheads. The integration times in Figures 14 & 15 are for the combination of the two bands. As per ALMA exposure calculator the

For these reasons, Scoville (2013) and Scoville et al. (2014) have proposed that a single frequency, broad-band measurement in the Rayleigh-Jeans tail of the dust SED (for instance at 345 GHz) is sufficient to establish dust and gas masses. Scoville and his colleagues argue that the variation of dust temperature on galactic scales is sufficiently small, such that the assumption of a constant dust temperature, $T_0 \sim 25$ K, is justified. Qualitatively this assumption is in very good agreement with the slow changes of average $T_{dust}$ with redshift and specific star formation rate in the stacked Herschel data (Magnelli et al. 2014) we presented in Figures 11 & 12. Based on SCUBA observations of a subset of the sample of z~0 disks from Draine et al. (2007), Scoville et al also argue that the dust to gas ratio does not vary significantly with metallicity ($\delta_{dg} \sim 0.0067 \sim$ const). This assumption is obviously at tension with the Leroy et al. (2011) scaling relation in equation (10) that predicts a fairly strong metallicity dependence of $\delta_{dg}$. This tension needs to be resolved in future studies.

The 'Rayleigh-Jeans tail' method thus is based on a single broad-band flux measurement, using calibrations of the 350 GHz dust emissivity to dust/gas mass from Planck observations in the Milky Way (Planck Collaboration 2011a,b) and of the nearby SINGS galaxies (Draine et al. 2007), yielding

$$\left(\frac{M_{molgas}}{1 \times 10^{10} M_\odot}\right) = \left(\frac{S_{350} D_L^2}{mJy \times Gpc^2}\right) \times (1+z)^{-(3+\beta)} \times \left(\frac{\nu_{obs}}{350\ GHz}\right)^{-(2+\beta)} \quad (26),$$

for $T_{dust} = 25$ K, the above calibration of $\kappa_{dust}$(350 GHz) (Scoville et al. 2014) and $\beta=1.8$.

It is instructive to compare this method to more detailed measurements that would include the temperature variations in Table 5, the full Planck formula of the dust modified grey-body emission, and the Leroy et al. (2011) recipe for $\delta_{dg}(Z)$ (equation (10)).

For this purpose we built a simulation to verify the performance of equation (26) on simulated data points with the above assumptions. We use the scaling relations in Tables 4 and 5 to define a grid of modified black body SEDs (MBBs) in the $M_*$-$sSFR$-$z$ space, using the binning by Magnelli et al. (2014). Each point of the grid is characterized by $M_*$, $sSFR$, $z$, $T_{dust}$ and $M_{molgas}$. Based on the scatter of our scaling relations we adopted $\sigma(logM_{molgas})=0.23$ and $\sigma(logT_{dust})=0.033$ to reflect variations in the average properties of a bin. We convolved the MBBs with a box filter centered on ALMA's band 7 (350 GHz), and computed the resulting flux density (in mJy). We then added Gaussian noise as computed for a given integration time from the ALMA sensitivity calculator assuming 34 12-meter antennas, given as the default in the cycle 2 time estimator. This resulting 'observed' flux density is then converted to $M_{molgas}$ using equation (26). Figure 13 (left panel) presents the result of the simulation and compares input to output gas masses. This Figure shows that the Rayleigh-Jeans approximation in equation (26), and assuming no metallicity dependent dust to gas ratio, leads to a **_significant underestimate of all_**

---

assumed water vapor column for the highest frequency band 9 is less than 0.47mm, for band 7 is less than 0.66mm and for band 6 is less than 1.3mm.

*inferred gas masses* (on average -0.3 dex), and to artificial *systematic trends throughout the probed parameter space* (±0.4 dex), and especially at high-z.

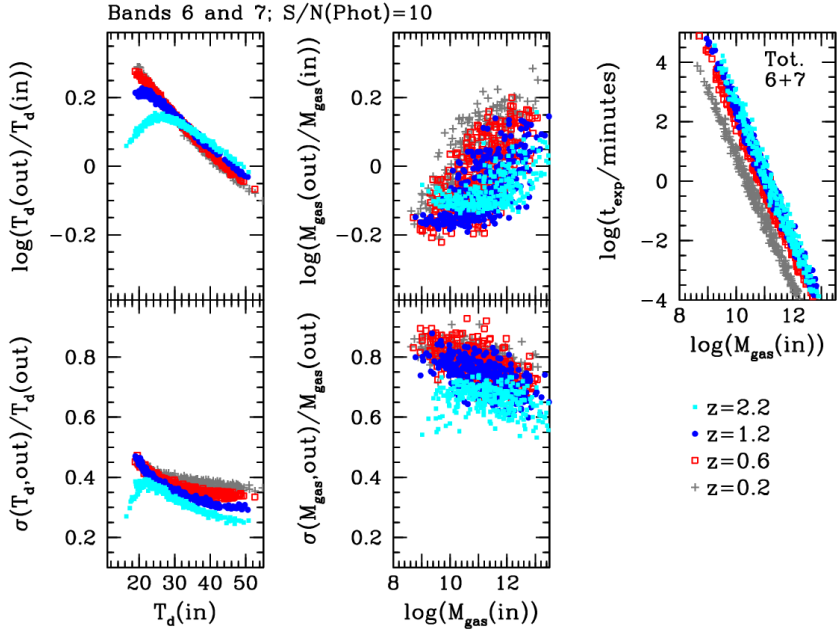

Figure 14: Performance of two band millimeter/submillimeter ALMA measurements (in bands 6 (240 GHz) and 7 (350 GHz)) for determining dust temperatures (left panels) and molecular gas mass (middle panels), as a function of input quantities for simulated galaxies on the Magnelli et al. (2014) grid, from the scaling relationships in Tables 4 and 5, and for 10σ photometry in each of the ALMA bands. The top panels show the logarithm of the ratio of the inferred quantity (dust temperature or gas mass) relative to the input quantity. The bottom panels show the 1σ fractional uncertainties in the temperature and gas mass estimates, given the flux density uncertainties of the measurements. The right panel gives the total ALMA integration time required assuming 34 antennas, for these two-band measurements, as a function of input gas mass. The color coding is for the different redshift bins as in Figures 2 & 5.

A first order improvement comes from introducing a constant Planck correction for all galaxies in any sample. For $T_{dust}=T_0=$ 25K=const one multiplies equation (26) with

$$\Gamma_{Pl} = \frac{\exp(h\nu_{obs}(1+z)/kT_0)-1}{h\nu_{obs}(1+z)/kT_0} \qquad (27),$$

where $\nu_{obs}$=350 GHz (Scoville et al. 2014). The central panel in Fig. 12 shows that with this global Planck correction the overall underestimate of gas masses is corrected and in fact over-compensated, mainly because the adopted value of $T_0$ is below the actual mean dust temperature near the main sequence. The large (±0.35 dex) systematic trends remain because of the intrinsic variation in $T_{dust}$ in Table 5 and the dust to gas ratio variations as a function of metallicity/mass. Correcting for the latter effect with the mass-metallicity relation in equation (10) improves the estimates further (right panel of Figure 13) but the temperature variations still cause significant systematic trends that should be accounted for in future studies attempting to measure fairly accurate relative trends in gas mass or depletion time. Obviously if the scaling relations in Table 5 are applicable to the galaxy sample a full Planck-correction with a variable $T_{dust}$ can be executed with equation (27), which should then give the correct output gas masses, making the single band technique indeed an efficient method for determining gas masses for large samples.

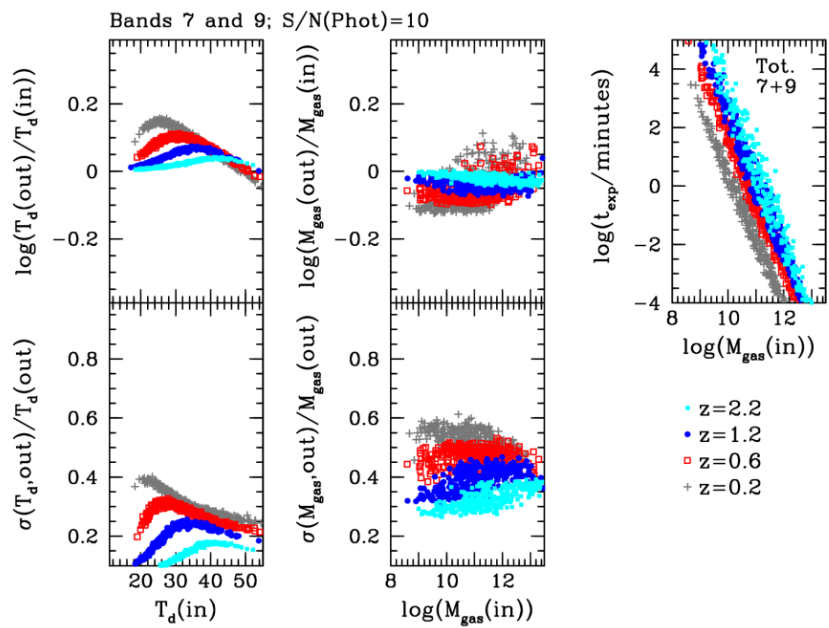

Figure 15: Performance of two band millimeter/submillimeter ALMA measurements (in bands 7 (350 GHz) and 9 (670 GHz)) for determining dust temperatures (left panels) and molecular gas mass (middle panels), as a function of input quantities for simulated galaxies on the Magnelli et al. (2014) grid, from the scaling relationships in Tables 4 and 5, and for 10σ photometry in each of the ALMA bands. The top panels show the logarithm of the ratio of the inferred quantity (dust temperature or gas mass) relative to the input quantity The bottom panels show the 1σ fractional uncertainties in the temperature and gas mass estimates, given the flux density uncertainties of the measurements. The right panel gives the total ALMA integration time required for these two-band measurements, as a function of input gas mass. The color coding is for the different redshift bins as in Figures 2 & 5.

If one does not wish to rely on the scaling relations in Table 5, a dust temperature must be established for every individual galaxy. This can be done from measurements in two ALMA bands, exploiting the frequency dependence of the Planck correction in equation (27). We have simulated the performance of such a two-band technique, starting with 10σ photometry in either the band 6 (240 GHz) and band 7 (350 GHz) combination (Figure 14), or the band 7 and band 9 (670 GHz) combination (Figure 15). The band 6/7 combination has the advantage of less demanding observing conditions but delivers less accurate output dust temperatures (left panels in Figure 14) and gas masses (middle panels in Figure 14). For 10σ photometry the resulting fractional precision of gas masses is not better than 0.7. The band 7/9 combination performs better in this respect (by about a factor of 2) but band 9 observations require much better, and thus somewhat rarer atmospheric conditions. For the quoted fractional precisions, the total observing times for an input gas mass of ~$10^{10}$ $M_\odot$ are >100 minutes at z~0.6-2.2. At least for z>1.5 this is still somewhat more favourable than CO detections (which are usually done in the 3mm atmospheric window) but the requirement of band 9 measurements for many galaxies likely will be quite restrictive in terms of sample size.

# 5. Conclusions

In this paper we have presented CO- and Herschel dust-based scaling relations of the molecular gas mass depletion time, of the molecular gas mass to stellar mass ratio, and of the dust temperature as a function of redshift, specific star formation rate offset and stellar mass, for each ~500 massive star forming galaxies (or stacks of star forming galaxies) between z~0 and 3, with a focus on the near main-sequence population. This is the first time that both dust and CO data have been compared in a large sample and on an equal footing across such a wide redshift range, and spanning 3 orders of magnitude in specific star formation rate and 1.8 orders of magnitude in stellar mass. In particular, the comparison of the CO- and dust-data allows us to derive quantitative constraints on the dependence of the CO conversion factor on redshift, specific star formation rate offset from the main-sequence, and stellar mass.

Our main results are

- in contrast to the rather controversial discussion in the past decades on possible large variations in the CO to molecular gas mass conversion factors, our study reveals a gratifying convergence of the CO- and dust-based analyses in the absolute zero points (absolute gas masses on the main sequence line as a function of redshift), and in the scaling indices with specific star formation rate offset and stellar mass, once the CO ladder excitation correction, and the metallicity dependence of the CO conversion factor and of the gas to dust mass ratio are taken into account. We emphasize, however, that this convergence only refers to massive SFGs with near-solar metallicity. The metallicity corrections used in this paper become large and probably unreliable at $<0.5\ Z_\odot$, corresponding to $M_\odot < 10^{10}\ M_\odot$.

   For massive SFGs and within ±0.6 dex of the main sequence, dust- and CO-based molecular gas masses agree to better than 50% throughout this large sampled range of parameters, and logarithmic scaling indices (power law exponents) agree to within fitting uncertainties of typically ±0.1. We show that this similarity sets stringent limits on changes of the CO conversion factor with redshift (less than a factor of 2 from z=0 to 2.5) and especially specific star formation rate (less than 25% across the main sequence).

   For outliers above the main sequence ($\Delta log(sSFR/sSFR(ms,z,M_*)) \geq 1$) the inferred CO conversion factor drops by a factor $\geq 2$, broadly consistent with earlier studies of the z~0 ULIRG population (Scoville et al. 1997, Downes & Solomon 1998, Bolatto et al. 2013), and implying a substantially higher radiation field density (Magdis et al. 2012b) and greater star formation efficiency than on the main-sequence. It is not clear whether this change is continuous or abrupt;
- depletion time scales on the main–sequence line change only slowly with redshift ($d(log t_{depl})/d(log(1+z))$= -0.34 (±0.15)), suggesting that the galactic scale star formation near the main-sequence is driven by similar physical processes across cosmic time, strengthening earlier findings by Tacconi et al. (2013) and Saintonge et al. (2013). As a result the ratio of molecular gas to stellar mass tracks the evolution of specific star formation rate, and both are plausibly controlled by the gas cycling into and out of galaxies (Magdis et al. 2012b, Tacconi et al. 2013, Santini et al. 2014, Sargent et al. 2014);

- depletion times change significantly with the ratio of the *sSFR* to that at the star formation main sequence in specific star formation rate (($d(logt_{depl})/d(log(sSFR/sSFR(ms,z,M*))$ = -0.5±0.01) and do not change much with stellar mass, in agreement with Saintonge et al. (2011), Sargent et al. (2014) and Huang & Kauffmann (2014). This in turn means that moving up in star formation rate at constant z and $M_*$ means increasing gas fractions and simultaneously lower depletion time scales, in about equal measure (Saintonge et al. 2012). The increase in 'star formation efficiency' with sSFR may be driven by internal parameters, such as the dense gas fraction (Lada et al. 2012) for the more compressed, cuspier SFGs above the main sequence (Wuyts et al. 2011b, Elbaz et al. 2011, Sargent et al. 2014);
- gas fractions drop with increasing stellar mass (see also Magdis et al. 2012b, Tacconi et al. 2013, Santini et al. 2014) because the star formation sequence flattens at the highest stellar masses near and above the quenching mass (the Schechter mass), plausibly as a result of feedback;
- because of the dependence of the depletion time scale on specific star formation rate (at a given redshift) observations of the galaxy integrated relation between molecular gas and star formation rate (the molecular KS-relation) in a sample of star forming galaxies naturally will exhibit a super-linear slope, although the intrinsic relation (at constant *sSFR*) is linear, in agreement with Kennicutt (et al. (2007), Bigiel et al. (2008), Genzel et al. (2010), Daddi et al. (2010b), Tacconi et al. (2013), Santini et al. (2014) and Sargent et al. (2014). The slope of the KS-relation can be anywhere between 1.0 and 1.7, shows substantial scatter for modest samples and steepens for an increasing range in specific star formation rate of the sample and increasing redshift;
- given that submillimeter detections of dust emission (e.g. with ALMA) are substantially more efficient than the detection of CO line emission, especially at $z \geq 1$, we have tested the reliability of single-frequency band continuum measurements of molecular gas masses across the parameter space sampled by our data. We find that without applying individual Planck-corrections in the dependence of continuum submillimeter flux densities on dust temperature, single band measurements will be affected by large systematic trends. These trends can be corrected for by the empirical scaling relation we have proposed here, or by two-band measurements. The latter require relatively long integrations and thus much of the initial advantage of the continuum technique (relative to CO-observations) is lost.

**Acknowledgements**

*We are grateful to the staff of the IRAM facilities for the continuing excellent support of the large CO survey programs we have been analysing in this paper. We also thank Albrecht Poglitsch and Matt Griffin, the PACS and SPIRE teams and ESA for their excellent work on the Herschel instruments and mission. We acknowledge Guinevere*

*Kauffmann for her leadership in bringing the COLDGASS program to fruition. We thank Alain Omont for his support of the PHIBSS programs at IRAM.*

Daddi, E., Elbaz, D., Walter, F., et al. 2010b, ApJL, 714, L118
Dale, D. A. & Helou, G. 2002, ApJ, 576, 159
Dame, T. M., Hartmann, D., & Thaddeus, P. 2001, ApJ, 547, 792
Dannerbauer, H., Daddi, E., Riechers, D. A., et al. 2009, ApJL, 698, L178
Danovich, M., Dekel, A., Hahn, O., Ceverino, D. & Primack, J.S. 2014, astro-ph 1407.7129
Davé, R., Finlator, K., & Oppenheimer, B. D., Fardal, M., Katz, N., Keres, D. & Weinberg, D.H. 2010 MNRAS, 404, 1355
Davé, R., Finlator, K., & Oppenheimer, B. D. 2011, MNRAS, 416, 1354
Davé, R., Finlator, K., & Oppenheimer, B. D. 2012, MNRAS, 421, 98
Davis, M., Guhathakurta, P., Konidaris, N. P., et al. 2007, ApJL, 660, L1
Dickman, R. L., Snell, R. L., & Schloerb, F. P. 1986, ApJ, 309, 326
Dobbs, C. L., Burkert, A. & Pringle, J. E. 2011, MNRAS, 413, 2935
Dobbs, C. L. & Pringle, J. E. 2013, MNRAS, 432, 653
Downes, D. & Solomon, P. M. 1998, ApJ, 507, 615
Draine, B. T., Dale, D. A., Bendo, G., et al. 2007 , ApJ, 663, 866
Draine, B. T. & Li, A. 2007, ApJ, 657, 810
Dutton, A.A., van den Bosch, F., Faber, S.M. et al. 2011, MNRAS 410, 1660
Elbaz, D., Daddi, E., Le Borgne, D., et al. 2007, A&A, 468, 33
Elbaz, D., Dickinson, M., Hwang, H. S., et al. 2011, A&A, 533, 119
Elmegreen, B.G. 1997, Rev.Mexicana Astron. Astrofis. Conf. Ser. 6, 165
Erb, D. K., Shapley, A. E., Pettini, M., et al. 2006, ApJ, 644, 813
Feldmann, R., Gnedin, N. Y. & Kravtsov, A. V. 2012a, ApJ, 747, 124
Feldmann, R., Gnedin, N. Y. & Kravtsov, A. V. 2012b, ApJ, 758, 127
Förster Schreiber, N. M., Genzel, R., Bouché, N., et al. 2009, ApJ, 706, 1364
Fu, J., Kauffmann, G., Li, C., & Guo, Q. 2012, MNRAS, 424, 2701
Galametz, M., Madden, S. C., Galliano, F., Hony, S., Bendo, G. J.& Sauvage, M. 2011, A&A, 532, 56
Gao, Y. & Solomon, P. M. 2004, ApJ, 606, 271
Genzel, R., Lutz, D., Sturm, E. et al. 1998, ApJ, 498, 579
Genzel, R., Tacconi, L. J., Gracia-Carpio, J., et al. 2010, MNRAS, 407, 2091
Genzel, R., Tacconi, L. J., Combes, F., et al. 2012, ApJ, 746, 69
Genzel, R., Förster Schreiber, N.M., Lang, P. et al. 2014, ApJ, 785, 75
Giavalisco, M., Ferguson, H. C., Koekemoer, A. M., et al. 2004, ApJ, 600, 93
Glover, S. C. O. & Clark, P. C. 2012, MNRAS, 421, 9
Garcia-Burillo, S., Usero, A., Alonso-Herrero, A., Gracia-Carpio, J., Pereira-Santaella, M, Colina, L., Planeseas, P. & Arribas, S. 2012, A&A, 539, 8
Gracia-Carpio, J., Garcia-Burillo, S. Planesas, P., Fuente, A. & Usero, A. 2008, A&A 479, 703
Gracia-Carpio, J. 2009, PhD thesis, University of Madrid
Gracia-Carpio, J. , Sturm, E. , Hailey-Dunsheath, S. et al. 2011, ApJ, 728, L7
Grenier, I. A., Casandjian, J.-M., & Terrier, R. 2005, Sci, 307, 1292
Greve, T., Bertoldi, F., Smail, I., et al. 2005, MNRAS, 359, 1165
Grogin,, N.A., Kocevski, D.D., Faber, S.M. et al. 2011, ApJS, 197, 35
Guilloteau, S., Delannoy, J., Downes, D., et al. 1992, A&A, 262, 624
53

**Table 1. CO sample**

| redshift range | N (±0.6 dex of ms) | N above 0.6 dex ms | N below 0.6 dex ms |
|---|---|---|---|
| 0 - 0.05 <>=0.033 N=296 | 193 | 54 | 49 (including 1 stack) |
| 0.05 - 0.45 <>=0.1 N=55 | 12 | 43 | 0 |
| 0.45 - 0.85 <>=0.67 N=48 | 30 | 18 | 0 |
| 0.85 – 1.2 <>= 1.1 N=32 | 26 | 5 | 1 |
| 1.2 – 1.75 <>=1.4 N=28 | 25 | 3 | 0 |
| 1.75 – 4.1 <>=2.3 N=41 | 28 | 11 | 2 |
| total 500 | 314 | 134 | 52 |

**Table 2. dust sample**

| mean redshift | N (±0.6 dex of ms) | N above 0.6 ms | N below 0.6 ms |
|---|---|---|---|
| 0.16  N=30 | 26 | 3 | 1 |
| 0.35  N=87 | 61 | 23 | 3 |
| 0.65  N=83 | 56 | 27 | 0 |
| 1     N=191 | 137 | 52 | 2 |
| 1.45  N=88 | 68 | 21 | 0 |
| 2     N=33 | 22 | 10 | 1 |
| total  N= 512 | 369 | 136 | 7 |

## Table 3. Parameters of power law fitting function for $t_{depl}$-scaling relations

|  | $a_{f1}$[a] | $\xi_{f1}$[a] | $\xi_{g1}$[a] | $\xi_{h1}$[a] |
|---|---|---|---|---|
| CO-data: binned | -0.04 (0.01) | -0.165 (0.04) | -0.46 (0.03) | -0.002 (0.03) |
| global | -0.025 (0.02) | -0.20 (0.06) | -0.43 (0.03) | -0.01 (0.03) |
| global, z=0 down-weighted[b] | -0.02 (0.024) | -0.16 (0.07) | -0.48 (0.03) | 0 (0.03) |
| dust-data: binned | 0.34 (0.07) | -0.77 (0.19) | -0.59 (0.05) | -0.01 (0.03) |
| global | 0.33 (0.07) | -0.74 (0.09) | -0.60 (0.03) | -0.00 (0.02) |
| average: binned | +0.07 (0.1) | -0.36 (0.1) | -0.51 (0.03) | -0.01 (0.02) |
| **global**[c] | **+0.1 (0.07)** | **-0.34 (0.05)** | **-0.49 (0.02)** | **+0.01 (0.03)** |
| global (Lilly)[d] | +0.01 (0.07) | -0.30 (0.05) | -0.5 (0.02) | -0.15 (0.02) |
| global ($g_1$(sSFR))[e] | -0.46 (0.07) | +1.18 (0.06) | -0.5 (0.02) | -0.197 (0.02) |
| global (FMR)[f] | +0.17 (0.07) | -0.45 (0.06) | -0.46 (0.02) | -0.02 (0.03) |

$$\log(t_{depl}(z, sSFR, M_*)|_{\alpha=\alpha_{MW}}) =$$
$$\log(f_1(z)|_{sSFR=sSFR(ms,z,M_*)})$$
$$+ \log(g_1(sSFR / sSFR(ms, z, M_*)))$$
$$+ \log(h_1(M_*))$$
$$= a_{f1} + \xi_{f1} \times \log(1+z) + \xi_{g1} \times \log(sSFR / sSFR(ms, z, M_*)) + \xi_{h1} \times (\log(M_*) - 10.5)$$

[a] in each of the columns the first fit value given comes from the 'parameter separated, binned' method discussed in sections 3.1 and 3.2 (6 redshift bins, first fitting the zero points of the $log(sSFR/sSFR(ms,z,M_*))$ dependence with an assumed slope of -0.46 (CO) and -0.59 (dust), then subtracting the zero points and fitting the $log(sSFR/sSFR(ms,z,M_*))$ slope for all data, then subtracting these fits values to finally fit the $logM_*$ dependence. All fits were made with power law functions. The second fit value comes from a direct, global fit to all data (not binned) in the 3-space, $log(1+z), log(sSFR/sSFR(ms,z,M_*), log\ t_{depl}$, and assuming no dependence on stellar mass (section 3.2.1), again with a linear fitting function. The values in parentheses are the 1σ fit uncertainties. In the case of the global fits these were determined by perturbing the original $log(sSFR/sSFR(ms,z,M_*))$ and $log\ t_{depl}$ measurements repeatedly with the ±0.2 dex errors and repeating the global fits. For the binned data main-sequence galaxies and main-sequence outliers (starbursts) ($log(sSFR/sSFR(ms,z,M_*))>0.6$) were given the same weight, while for the global fits main-sequence galaxies were given 60% greater weight than the outliers to take into account the lower relative uncertainties in the determination of their stellar masses and star formation rates.

[b] to explore the influence of the unequal numbers of points (~300 z≤0.05 data from COLDGASS and GOALS) we down-weighted each of these by 2.7 to give all the z-bins approximately the same weight.

[c] For the global combined CO+dust fit we first added 0.1 dex to all CO depletion time values, and likewise subtracted 0.1 dex for all dust depletion times before carrying out the global fit, in order to bring the two data sets to the same zero-point.

[d] For this row we carried out a combined CO and dust global fit (1012 data points) we proceeded as above in [b] but now employed the Lilly et al. (2013) prescription for the

main-sequence, $sSFR(ms,z,M_*)=0.117\ (Gyr^{-1}) \times (M_*/3.16\times10^{10}M_\odot)^{-0.1} \times (1+z)^3$ for z<2, and $sSFR(ms,z,M_*)=0.5\ (Gyr^{-1}) \times (M_*/3.16\times10^{10}M_\odot)^{-0.1} \times (1+z)^{1.667}$ for z≥2, instead of the one by Whitaker et al. (2012, eq.1). The Lilly et al. (2013) fitting function is a simple power law in stellar mass (without curvature, as in Whitaker et al.). It captures the actual location of the SFGs in the stellar mass- *sSFR* plane quite well at $logM_*$=9.5 to 10.5 and z=0=2.5 but the more massive galaxies then systematically lie below the Lilly et al. fit.
[e]For this row we again combined the CO and dust data in a global fit and assumed that $g_1$ depends only on *sSFR* ($Gyr^{-1}$), without referring to the main sequence.
[f]Here we have replaced the estimation of metallicities from equation (12) (mass-metallicity relation) to the fundamental metallicity relation of Mannucci et al. (2010), involving stellar mass *and* star formation rate (equation (12a)). As a result, metallicities drop and the conversion factor in equation (8) is greater than for equation (12).

**Table 4. parameters of power law fitting function for $M_{molgas}/M_*$-scaling relations**

|  | $a_{f2}{}^a$ | $\xi_{f2}{}^a$ | $\xi_{g2}{}^a$ | $\xi_{h2}{}^a$ |
|---|---|---|---|---|
| CO-data | -1.23 (0.01) | +2.71 (0.09) | +0.51 (0.03) | -0.35 (0.03) |
| global | -1.12 (0.012) | +2.71 (0.06) | +0.53 (0.02) | -0.35 (0.02) |
| dust-data | -0.87 (0.06) | +2.26 (0.24) | +0.51 (0.05) | -0.41 (0.03) |
| global | -0.98 (0.03) | +2.32 (0.1) | +0.36 (0.04) | -0.40 (0.04) |
| average | -1.1 (0.05) | +2.6 (0.1) | +0.51 (0.03) | -0.38 (0.03) |
| **global[b]** | **-1.11 (0.02)** | **+2.68 (0.05)** | **+0.49 (0.03)** | **-0.37 (0.04)** |
| global (Lilly)[c] | -0.98 (0.02) | +2.65 (0.05) | +0.50 (0.03) | -0.25 (0.03) |
| global (sSFR)[d] | -0.51 (0.02) | +1.18 (0.06) | +0.50 (0.03) | -0.198 (0.03) |
| global (FMR)[e] | -1.05 (0.02) | +2.6 (0.06) | +0.54 (0.03) | -0.41 (0.04) |

$$\log(M_{molgas}/M_*(z, sSFR, M_*)|_{\alpha=\alpha_{MW}}) =$$
$$\log(f_2(z)|_{sSFR=sSFR(ms,z,M_*)})$$
$$+\log(g_2(sSFR/sSFR(ms,z,M_*)))$$
$$+\log(h_2(M_*))$$
$$= a_{f2} + \xi_{f2} \times \log(1+z) + \xi_{g2} \times \log(sSFR/sSFR(ms,z,M_*)) + \xi_{h2} \times (\log(M_*)-10.77)$$

[a] in each of the columns the first fit value given comes from the 'parameter separated, binned' method discussed in sections 3.1 and 3.2 (6 redshift bins, first fitting the zero points of the $log(sSFR/sSFR(ms,z,M_*))$ dependence with an assumed slope of 2.7 (CO) and -2.6 (dust), then subtracting the zero points and fitting the $log(sSFR/sSFR(ms,z,M_*))$ slope for all data, then subtracting these fits values to finally fit the $logM_*$ dependence. All fits were made with power law functions. The second fit value comes from a direct, global fit to all data (not binned) in the 4-space, $log(1+z)$, $log(sSFR/sSFR(ms,z,M_*)), log(M_*), log(M_{gas}/M_*)$, again with a linear fitting function. The values in parentheses are the 1σ fit uncertainties. In the case of the global fits these were determined by perturbing the original $log(sSFR/sSFR(ms,z,M_*))$ and $log(M_{gas}/M_*)$, measurements repeatedly with the ±0.2 dex errors, and the $log(M_*)$ values with ±0.15 dex errors, and repeating the global fits. For the binned data main-sequence galaxies and main-sequence outliers (starbursts) ($log(sSFR/sSFR(ms,z,M_*))>0.6$) were given the same weight, while for the global fits main-sequence galaxies were given 60% greater weight than the outliers to take into account the lower relative uncertainties in the determination of their stellar masses and star formation rates.

[b] For the global combined CO+dust fit (1012 data points) we first added 0.1 dex to all CO $M_{gas}/M_*$ values, and likewise subtracted 0.1 dex for all dust $M_{gas}/M_*$ values before carrying out the global fit, in order to bring the two data sets to the same zero-point.

[c] For this row we carried out a combined CO and dust global fit (1012 data points) we proceeded as above in [b] but now employed the Lilly et al. (2013) prescription for the main-sequence, $sSFR(ms,z,M_*)=0.117 (Gyr^{-1}) \times (M_*/3.16\times10^{10}M_\odot)^{-0.1} \times (1+z)^3$ for z<2, and $sSFR(ms,z,M_*)=0.5 (Gyr^{-1}) \times (M_*/3.16\times10^{10}M_\odot)^{-0.1} \times (1+z)^{1.667}$ for z≥2, instead of the one by Whitaker et al. (2012, eq.1). The Lilly et al. (2013) fitting function is a simple power law in stellar mass (without curvature, as in Whitaker et al.). It captures the actual location of the SFGs in the stellar mass- sSFR plane quite well at $logM_*$=9.5 to 10.5 and z=0=2.5 but the more massive galaxies then systematically lie below the Lilly et al. fit.

[d] For this row we again combined the CO and dust data in a global fit and assumed that $g_2$ is only a function of sSFR ($Gyr^{-1}$), without referring to the main sequence.

[e] Here we have replaced the estimation of metallicities from equation (12) (mass-metallicity relation) to the fundamental metallicity relation of Mannucci et al. (2010), involving stellar mass **and** star formation rate (equation (12a)). As a result, metallicities drop and the conversion factor in equation (8) is greater than for equation (12).

## Table 5. Parameters of power law fitting function for $T_{dust}$-scaling relations

| | $a_{f3}$ | $\xi_{f3}$ | $\xi_{g3}$ | $\xi_{h3}$ |
|---|---|---|---|---|
| dust-data | 1.432 (0.006) | +0.105 (0.02) | +0.086 (0.003) | -0.012 (0.003) |

$$\log(T_{dust}) = \log(f_3(z)|_{sSFR=sSFR(ms,z)}) + \log(g_3(sSFR/sSFR(ms,z))) + \log(h_3(M_*))$$
$$= a_{f3} + \xi_{f3} \times \log(1+z) + \xi_{g3} \times \log(sSFR/sSFR(ms,z)) +$$
$$\xi_{h3} \times (\log(M_*)-10.5)$$